\documentclass[prb,twocolumn]{revtex4-2}
\usepackage[a4paper,top=2.00cm,bottom=2.00cm,left=2.00cm,right=2.00cm]{geometry}
\usepackage{amsmath}
\usepackage{amssymb}
\usepackage{enumerate}
\usepackage[force]{feynmp-auto}
\usepackage{natbib}
\setcitestyle{super}
\usepackage[utf8]{inputenc}
\fontfamily{cmr}\selectfont
\usepackage{graphicx}
\DeclareGraphicsExtensions{.eps, .pdf, .jpg, .png}
\usepackage[colorlinks=true,urlcolor=blue,linkcolor=blue,citecolor=blue]{hyperref}
\usepackage{bookmath}
\begin{document}
\title{Pair Fluctuation Effects on Quasiparticle Transport in Fermi Systems}
\author{Wei-Ting Lin}
\author{J. A. Sauls}
\affiliation{Department of Physics \& Northwestern-Fermilab Center for Applied Physics and 
             Superconducting Technologies, Northwestern University, 
             Evanston, Illinois 60208, USA}
\date{\today}
\begin{abstract}
The leading corrections to Fermi liquid theory for non-equilibrium quasiparticle transport near a Cooper instability arise from the virtual emission and absorption of incipient Cooper pairs. We formulate the corrections to the Landau-Boltzmann transport equation starting from Keldysh's field theory for non-equilibrium, strongly interacting Fermions.
The theory is applicable to quasiparticle transport in conventional and unconventional superconductors, dense nuclear matter and the low temperature phases of liquid \He.
Validation of the theory is provided by our analysis, and quantitative agreement between theory and experiment, of the excess attenuation of zero sound in liquid \He\ near the superfluid transition. We propose an additional experimental test of the theory based on the effects of a Zeeman field on the spectrum of pairing fluctuations for the Cooper instability in spin-triplet superfluids.
\end{abstract}
\maketitle
\noindent{\it Introduction --}
Close to a normal to superconducting transition incipient Cooper pairs, ``pair fluctuations'', become sufficiently long-lived that they lead to observable corrections to transport processes.
A well studied example is ``paraconductivity''. The decrease in resistivity as $T\rightarrow T_c$ from above the transition is attributed to enhanced conductivity from pair fluctuations.~\cite{glo67,sko75,asl68a,mak68,tho70}
Pair fluctuations are also important for understanding the BCS-BEC crossover which can be realized in Fermionic cold gases,~\cite{ens11,ens12,liu14} and are believed to play a role in understanding the pseudogap in high-temperature superconductors.~\cite{esc99a,che05a}

Low temperature transport processes in Fermi liquids are governed by a kinetic theory of Fermionic quasiparticles, i.e. long-lived particle and hole excitations of the Fermionic vacuum. Their distribution function obeys the Boltzmann-Landau transport equation, with collision processes arising from other quasiparticles, impurities or phonons.~\cite{lan57,abr57}
In Fermi liquids that are close to a second-order phase transition quasiparticle scattering by damped, but persistent, fluctuations of the incipient order parameter can be significant. A good example is a Fermi liquid with strong short-range repulsive interactions. Repeated scattering of spin-triplet particle-hole pairs leads to long-lived ``paramagnons'', i.e. a damped, but slowly decaying ferromagnetic magnons.~\cite{lev83} 
These Bosonic fluctuations open a new channel for quasiparticle-paramagnon scattering, which has been studied extensively in the context of nearly ferromagnetic Fermi liquids.
This is a ``feedback'' effect which leads to non-analytic corrections to the Fermi liquid predictions for the heat capacity and transport coefficients,~\cite{don66,pet69} and can also mediate spin-triplet Cooper pairing.~\cite{lay71,lay74} Indeed this is the widely accepted mechanism believed to be responsible for the spin-triplet, p-wave superfluid phases of \He.~\cite{and75,leg75,vollhardt90}

Here we report new results for the corrections to the Boltzmann-Landau transport equation derived from the quasiparticle self-energy for emission and absorption of virtual pair fluctuations. 
Starting from a Keldysh formulation of the Dyson equations for the non-equilibrium propagators of an interacting Fermion system we derive the leading order corrections to the collision integral and molecular field self-energy resulting from the interaction of quasiparticles with damped, but long-lived pair fluctuations for the normal-state Landau-Boltzmann transport equation.
The theory is broadly applicable to non-equilibrium quasiparticle transport near the Cooper instability in superconductors or superfluids. 

Compared to the superconductors, or even cold atoms, and in spite of the fact that both the normal and superfluid phases of liquid \He\ have been studied extensively,~\cite{vollhardt90, dobbs00} there is limited research on the border between the two phases where pair fluctuations become strong. 
As an application of our theoretical formulation we consider zero sound propagation and attenuation near the superfluid transition. Zero sound in \He\ exists at low temperature as a coherent particle-hole excitation of the Fermi surface, weakly damped by scattering off thermal quasiparticles. Near the superfluid transition particle-hole excitations comprising zero sound can also scatter off pair fluctuations leading to anomalies in the attenuation and velocity.

%
\noindent{\it Keldysh Transport Equations --}
Our theory is based on a quantum transport equation derived from Keldysh's formulation of non-equilibrium field theory.~\cite{kel65} This approach differs from the theory for pairing fluctuation effects in superconductors based on the Kubo and Matsubara formalism, and focuses on transport by pair fluctuations.~\cite{asl68a,mak68,tho70} In particular, the theory we present highlights the effects of pair fluctuations on nonequilibrium \emph{quasiparticle transport} near $T_c$. Below we summarize the key elements of the corrections to the Boltzmann-Landau transport equation for a Fermi liquid near a Cooper instability. Details and supporting analysis of this theory are presented in a separate report in Ref.~\onlinecite{lin21}.

The quantum transport equation obtained from the Dyson equation for the Keldysh path-ordered Green's function is~\cite{lin21}
\vspace*{-3mm}
\begin{eqnarray}
\partial_t n_{\vp} 
&+&
\nicefrac{1}{m}\vp\cdot\grad_{\vR}n_{\vp}
+
\grad_{\vp}\mathfrak{Re}\Sigma^{11}_{p}\cdot\grad_{\vR}n_{\vp}
\nonumber\\ 
&-&
\grad_{\vR}\mathfrak{Re}\Sigma^{11}_{p}\cdot\grad_{\vp}\,n_{\vp}
-
\partial_{\eps}\mathfrak{Re}\Sigma^{11}_{p}\big|_{\eps=\eps_{\vp}}\partial_t n_{\vp}
\nonumber\\
&=&
-i
\left(\Sigma^{21}_{p}\,n_{\vp}
+
\Sigma^{12}_{p}\,(1-n_{\vp})
\right)
\,,
\label{eq-Keldysh_transport_equation}
\vspace*{-3mm}
\end{eqnarray}
where $p = (\vp,\eps)$ is the four-momentum. 

\begin{widetext}
\begin{figure}[t]
\begin{minipage}{\textwidth}
\centering
\begin{fmffile}{fmf-Self_Energy_High}
  \begin{fmfgraph*}(100,45)
    \fmfpen{thin}
    \fmfset{arrow_len}{3mm}
    \fmftop{b1}
    \fmfv{label=(a) $\sml^0\hspace*{25mm}$}{b1}
    \fmfleft{l1,l2}
    \fmfright{r1,r2}
    \fmfv{label=$a$,l.d=0.5mm,l.a=180}{l1}
    \fmfv{label=$b$,l.d=0.5mm,l.a=0}{r1}
    \fmfv{decor.shape=circle,d.f=0,d.si=8mm,label=$\Sigma_{p}^{\mbox{\tiny high}}$,l.d=-1.5mm}{v1}
    \fmf{fermion,tension=.9,label=$p$,l.a=+90}{r1,v1}
    \fmf{fermion,tension=.9,label=$p$,l.a=-90}{v1,l1}
  \end{fmfgraph*}
\end{fmffile}
\begin{fmffile}{fmf-Self_Energy_Landau}
  \begin{fmfgraph*}(100,45)
    \fmfpen{thin}
    \fmfset{arrow_len}{3mm}
    \fmftop{b1}
    \fmfv{label=(b) $\sml^1\hspace*{25mm}$}{b1}
    \fmfleft{l1,l2}
    \fmfright{r1,r2}
    \fmfpolyn{empty,tension=.2,label=$\Gamma^{\mathsf{ph}}$}{G}{4}
    \fmf{fermion,tension=.9,label=$p$,l.a=+90}{r1,G1}
    \fmf{fermion,right,tension=0.1,label=$p'$}{G2,G3}
    \fmf{fermion,tension=.9,label=$p$,l.a=-90}{G4,l1}
    \fmfv{label=$a$,l.a=180,l.d=0.5mm}{l1}
    \fmfv{label=$b$,l.d=0.5mm,l.a=0}{r1}
  \end{fmfgraph*}
\end{fmffile}
\begin{fmffile}{fmf-Self_Energy_Binary}
  \begin{fmfgraph*}(120,45)
    \fmfpen{thin}
    \fmfset{arrow_len}{3mm}
    \fmftop{b1}
    \fmfv{label=(c) $\sml^2\hspace*{25mm}$}{b1}
    \fmfleft{l1,l2}
    \fmfright{r1,r2}
    \fmfv{label=$a$,l.d=0.5mm,l.a=180}{l1}
    \fmfv{label=$b$,l.d=0.5mm,l.a=0}{r1}
    \fmfpolyn{empty,tension=0.2,label=$\Gamma$}{G}{4}
    \fmfpolyn{empty,tension=0.2,label=$\Gamma$}{H}{4}
    \fmf{fermion,tension=0.9,label=$p$,l.a=90}{r1,H1}
    \fmf{fermion,tension=0.9,label=$p'''$,l.d=-4mm}{H4,G1}
    \fmf{fermion,tension=0.0,label=$p''$,l.a=90}{G2,H3}
    \fmf{fermion,right,tension=0.0,label=$p'$,l.a=90}{H2,G3}
    \fmf{fermion,tension=0.9,label=$p$,l.a=90}{G4,l1}
  \end{fmfgraph*}
\end{fmffile}
\begin{fmffile}{fmf-Self_Energy_Cooper}
  \begin{fmfgraph*}(100,45)
    \fmfpen{thin}
    \fmfset{arrow_len}{3mm}
    \fmftop{b1}
    \fmfv{label=(d) $\sml^3\hspace*{25mm}$}{b1}
    \fmfleft{l1,l2}
    \fmfright{r1,r2}
    \fmfv{label=$a$,l.d=0.5mm,l.a=180}{l1}
    \fmfv{label=$b$,l.d=0.5mm,l.a=0}{r1}
    \fmfpolyn{empty,tension=.2,label=$\Gamma_{Q}$}{G}{4}
    \fmf{fermion,tension=.9,label=$p$,l.a=90}{r1,G1}
    \fmf{fermion,left,tension=0.1,label=$-p+Q$}{G3,G2}
    \fmf{fermion,tension=.9,label=$p$,l.a=90}{G4,l1}
  \end{fmfgraph*}
\end{fmffile}
\caption{Self energy diagrams and their order of magnitude in the small expansion 
         parameter, $\sml\in\{\kb T_c/E_f,1/k_f\xi\}$ of Fermi liquid theory.
(a) $\Sigma_{p}^{\mbox{\tiny high}}$ is $\cO(\sml^0)$ and gives the Fermi liquid mass 
    renormalization,
(b) is the Landau mean field self energy of $\cO(\sml^1)$,
(c) is the self energy from binary collision scattering and is $\cO(\sml^2)$, and
(d) is the self energy derived scattering of quasiparticles by pair fluctuations and 
    is formally $\cO(\sml^3)$, but the vertex $\Gamma_Q$ diverges for 
       $T\rightarrow T_c^+$ and $Q\rightarrow 0$.
}
\label{fmf-Self_Energies}
\end{minipage}
\end{figure}
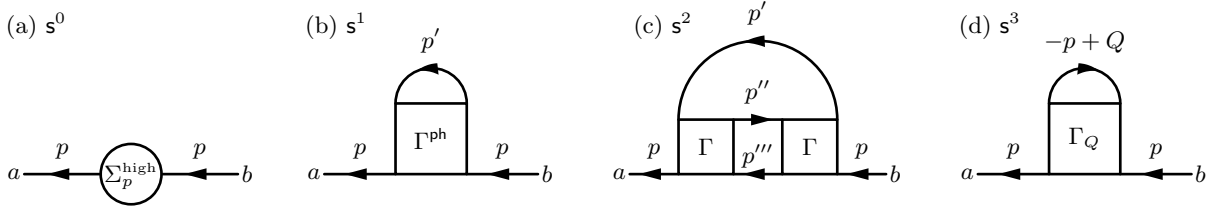

\end{widetext}

Equation~\eqref{eq-Keldysh_transport_equation} is the starting point for deriving the Boltzmann-Landau transport equation for a quasiparticle distribution function, $n_{\vp}(\vR,t)\equiv -i\int d\eps\,G^{12}_{p}(\vR,t)$, evolving in phase space $(\vp,\vR)$.
The first two terms in Eq.~\eqref{eq-Keldysh_transport_equation} describe the space-time evolution of $n_{\vp}$ for \emph{non-interacting} Fermions of mass $m$.
Many particle interactions are encoded in the self-energy functional, $\Sigma^{ab}_{p}(\vR,t)$, where $a,b\in\{1,2\}$ are the Keldysh indices for the path-ordered Keldysh Green's function and self-energy; in particular $a=1$ ($a=2$) represents the forward (backward) branch of the Keldysh path-ordered contour.~\cite{lin21,kit10} The diagonal component, $\Sigma^{11}_p$, contributes to the smooth space-time evolution of the distribution function on the left side of Eq.\eqref{eq-Keldysh_transport_equation}, while the off-diagonal term with $\Sigma^{12}_p$ ($\Sigma^{21}_p$) reduces to the ``scattering in'' (``scattering out'') contribution to the collision integral. 

For low temperatures, low excitation energies, long wavelength disturbances, and weak disorder, we can classify the contributions to the self-energy in an asymptotic expansion in the small ratios, $\sml\in\{\kb T/E_f, \hbar\omega/E_f,\hbar q/p_f, \hbar/\tau E_f\,\ldots\}$, characteristic of strongly interacting Fermi liquids.~\cite{ser83,rai94b} 
The zeroth order self-energy (Fig.~\ref{fmf-Self_Energies}(a)), $\Sigma_{p}^{\mbox{\tiny high}}$, renormalizes the Fermi velocity, i.e. $\vp/m\rightarrow \vv_{\vp}=\vp/m+\grad_{\vp}\Sigma_{p}|_{p=p_f}=p_f/m^*\hat\vp$, where $m^*$ is the quasiparticle effective mass.\cite{AGD}

Fig.~\ref{fmf-Self_Energies}(b) represents the leading order correction to $\Sigma^{11}_{\vp}$ from particle-hole scattering in the forward direction, i.e. the Landau channel. This term gives the mean-field correction to the quasiparticle energy that depends on the \emph{deviation} of the distribution function, $\delta n_{\vp}$, from its equilibrium form, 
\vspace*{-2mm}
\begin{equation}
\hspace*{-2mm}
\Sigma^{\text{MF}}_{\vp} 
\ns=\ns\ns\sum_{\vp,\sigma}f^{s}(\vp,\vp')\delta n_{\vp'}
\ns=\ns\ns\int\ns dS_{\hat\vp'}F^{s}(\hat\vp\cdot\hat\vp')\nu_{\hat\vp'}
\,,
\vspace*{-1mm}
\end{equation}
where the distribution function is expressed as a deformation of the Fermi surface, $\nu_{\hat\vp} \equiv \int d\eps_{\vp}\,\delta n_{\vp}$, $F^s(\hat\vp\cdot\hat\vp')=2N(0)\Gamma^{\mathsf{ph}}(p_f\hat\vp,0;p_f\hat\vp',0)$ is the dimensionless Landau interaction, $2N(0)=m^* p_f/\pi^2\hbar^3$ is the density of states at the Fermi energy, and the integration is over the Fermi surface $S_{\hat\vp}$.\footnote{Here we consider non-magnetic disturbances of the quasiparticle distribution in which case only the spin-independent Landau interaction is relevant and the sum over spin states contributes a factor of $2$.}

The self-energy diagram in Fig.~\ref{fmf-Self_Energies}(c) derives from binary collisions of quasiparticles, is of order $\sml^2$ due to phase space restrictions from Fermi statistics, and contributes the leading order contribution to the collision integral,
\vspace*{-3mm}
\begin{equation}
I_{\vp}[n_p]\equiv-i\left(\Sigma^{(c)21}_{\vp}n_{\vp}+\Sigma^{(c)12}_{\vp}(1-n_{\vp})\right) \,.
\label{eq-collision_integral1}
\end{equation}
To $\cO(\sml^2)$ the collision integral reduces to a functional of the quasiparticle distribution function for excitations with momenta and energy and near the Fermi surface,
\begin{eqnarray}
&\hspace*{-2.25mm}
I_{\vp}[n_p]&
\hspace*{-1.0mm}
=
-\ns\ns\ns\sum_{p_2,p_3,p_4}\ns W_{\text{N}}(p,p_2;p_3,p_4)\,\delta_{\sum_i \vp_i}\delta_{\sum_i\eps_{p_i}}\delta_{\sum_i\sigma_i}
\hspace*{5mm}
\label{eq-collision_integral2}
\\
&\times\Big[n_p& 
\hspace*{-0.0mm}
n_{p_2}(1\ns-\ns n_{p_3})(1\ns-\ns n_{p4})-(1\ns-\ns n_p)(1\ns-\ns n_{p_2})n_{p_3} n_{p4}\Big]
,
\nonumber
\end{eqnarray}
where $W_{\text{N}}(p_1,p_2;p_3,p_4)\equiv 2\pi|\mathsf{T}(p_1,p_2;p_3,p_4)|^2$ is the transition rate for binary collision scattering of quasiparticles, $(p_1,p_2)\rightarrow(p_3,p_4)$. The delta functions in Eq.~\eqref{eq-collision_integral2} enforce energy, momentum, spin conservation in collisions. 

Equation~\eqref{eq-Keldysh_transport_equation} then reduces to the Boltzmann-Landau transport equation 
\begin{equation}
\pder{}{t}n_{\vp}
+\vv_{\vp}\cdot\grad_{\vR}n_{\vp}
-\grad_{\vR}\Sigma^{\mbox{\tiny MF}}_{\vp}\cdot\grad_{\vp}n_{\vp}
= I_{\vp}[n_p]
\,,
\label{eq-Boltzmann-Landau_equation}
\end{equation}
which provides the theoretical framework of nonequilibrium transport processes in strongly interacting Fermi liquids, and which have been studied extensively.~\cite{bro70,abrikosov78,bay78,sau10} In particular, the binary collision scattering described by Eq.~\eqref{eq-collision_integral2} leads to a mean quasiparticle scattering rate given by $\frac{1}{\tau_{\text{N}}}=\nicefrac{\pi^2}{4}\,N(0)^2\,\langle W_{\text{N}}\rangle\,(\kb T)^2/E_f$, where $\langle W_{\text{N}}\rangle$ is an average of the scattering probability over the Fermi surface.~\cite{sau81b} This sets the time scale for inelastic quasiparticle collisions and determines the signature scaling behavior in clean Fermi liquids: for viscosity, $\eta\propto T^{-2}$, thermal conductivity, $\kappa \propto T^{-1}$ and zero sound attenuation, $\alpha_0\propto T^2$.\cite{bay78}

The results of this report derive from the process represented by Fig.~\ref{fmf-Self_Energies}(d), which is the leading correction to Fermi liquid theory for a non-equilibrium distribution of quasiparticles scattering off pair fluctuations. The scattering amplitude, $\Gamma_Q$, is the sum of repeated scatterings of nearly zero-momentum, zero-energy $(Q\rightarrow 0)$ quasiparticle pairs. This amplitude becomes important when the temperature of the Fermi system approaches the pairing transition temperature, at which point $\Gamma_Q$ diverges.
The Keldysh components of this amplitude satisfy a Bethe-Salpeter equation in the particle-particle (Cooper) channel,
\begin{eqnarray}
\hspace{-1em}
&
\Gamma_{\alpha\beta,\gamma\delta}^{ab}(p,p';Q) 
&
=
\Gamma^{\mathsf{pp}}_{\alpha\beta,\gamma\delta}(p,p')
\check{\tau}_3^{ab} 
+
\Gamma^{\mathsf{pp}}_{\alpha\beta,\alpha'\beta'}(p,p'')
\nonumber
\\ 
&
\hspace*{-13mm}
\times
&
\hspace*{-15mm}
\check{\tau}_3^{ac} 
G^{cd}(p'')G^{cd}(Q-p'')
\Gamma_{\alpha'\beta',\gamma\delta}^{db}(p'', p';Q)
\,,
\label{eq-Bethe-Salpeter_Cooper_Channel}
\end{eqnarray}
where $\Gamma^{\mathsf{pp}}(p,p')$ is the two-particle irreducible vertex, i.e. the ``pairing interaction'', responsible for Cooper instability. The pair of Greek indices $\alpha\beta$ ($\gamma\delta$) are spin projections of the outgoing (incoming) pairs, while the Latin indices, $ab$, are Keldysh path indices. Note that $\check\tau_3$ is the diagonal two-component Pauli matrix in Keldysh space, and the normal-state Keldysh propagator, $G^{cd}(p)$, is independent of spin in zero magnetic field.

We evaluate the scattering amplitude and the self-energy using the Keldysh formalism and obtain corrections to (i) the Landau mean field self-energy from $\mbox{Re}\Sigma^{(d)11}_{\vp}$ and (ii) the collision integral from $\Sigma^{(d)12}_{\vp}$ and $\Sigma^{(d)21}_{\vp}$.
In particular, we obtain the correction to the collision integral from quasiparticle-pair-fluctuation scattering with scattering probability~\cite{lin21} 
\vspace*{-3mm}
\begin{eqnarray}
&
\displaystyle{
W_{\text{C}}(\vp,\vp';\vQ,\omega) 
= 
\pi \Big|\frac{4\pi}{N(0)}\Big|^2
\,\times
(2S+1)
}
\hspace*{10mm}
&
\label{eq-Scattering_probability_QP-Pair}
\\
&
\displaystyle{
\times
\sum_{m=-l}^{+l}
\bigg| 
         Y_{lm}(\hat{\vp})
\frac{1}{\vartheta+\xi_{lm}^2 Q^2-i\frac{\pi\omega}{8T}} 
         Y_{lm}^*(\hat{\vp}')
\bigg|^2
}
\,,
&
\nonumber
\vspace*{-3mm}
\end{eqnarray}
for the scattering of a quasiparticle with momentum $\vp$ to $\vp'$ by pair fluctuations with center of mass momentum, $\vQ$, and energy, $\omega$. The quantity $\vartheta\equiv(T-T_{\mathrm{c}})/T_{\mathrm{c}}$ is the reduced temperature.
The factor 
\vspace*{-3mm}
\begin{equation}
\cC_{lm}(Q,\omega)=\frac{1}{\vartheta + \xi_{lm}^2 Q^2 - i\pi\omega/8T}
\,,
\label{eq-Cooper_Pair_propagator}
\vspace*{-2mm}
\end{equation}
is the propagator for Cooper pair fluctuations of spin $S$ and relative orbital angular momentum $l,m$ which has a pole at $Q\equiv(\vQ,\omega)=0$ for $T=T_{c}$. The correlation lengths are all of order, $\xi_{lm}\propto\xi_0\equiv\hbar v_f/2\pi T_c$. In the absence of a magnetic field there are $(2S+1)(2l+1)$ pair fluctuations contributing to the scattering probability.
Equation~\eqref{eq-Scattering_probability_QP-Pair} is easily generalized to unconventional superconductors with point group symmetry by replacing the sum over spherical harmonics by a sum over the basis functions of the irreducible representation corresponding to the pairing channel.

%
\noindent{\it Zero Sound \& Pair Fluctuations --}
As an application of the theory we consider the corrections to the attenuation and velocity of zero sound from the scattering of quasiparticles by pair fluctuations in normal \He\ near the superfluid transition. 
We compare our theoretical prediction with measurements of Paulson and Wheatley,\cite{pau78c} as well as our theoretical result for the quasiparticle-pair fluctuation contribution to the collision integral with the phenomenological theory of Emery,\cite{eme76} and Samalam and Serene.\cite{sam78}

Zero sound is a propagating mode of liquid \He\ that exists at low temperature as a coherent particle-hole excitation of the Fermi surface, weakly damped by scattering off thermal quasiparticles.
Undamped zero sound is an eigenmode of the collisionless (i.e. $I_{\vp}=0$) transport equation (Eq.~\eqref{eq-Boltzmann-Landau_equation}). Small deviations of the distribution function from the Fermi distribution, $n_0(\varepsilon_{\vp})$, can be expressed as a deformation of the Fermi surface, $\delta n_{\vp}=(-\partial{n_0(\varepsilon_{\vp})}/\partial\varepsilon_{\vp})\,\nu_{\hat\vp}(\vr,t)$. The Fourier modes, $\nu_{\hat\vp}(\vk,\omega)$, obey the equation, 
\vspace*{-3mm}
\begin{equation}
(\omega -\vv_{\vp}\cdot\vk)\nu_{\hat\vp}
        -\vv_{\hat\vp}\cdot\vk\,\int dS_{\hat\vp'}F^s(\hat\vp\cdot\hat\vp')\nu_{\hat\vp'}=0
\,.
\label{eq-collisionless_sound}
\vspace*{-3mm}
\end{equation}

\begin{figure}[t]
\vspace{-0.2em}
\includegraphics[width=0.5\textwidth]{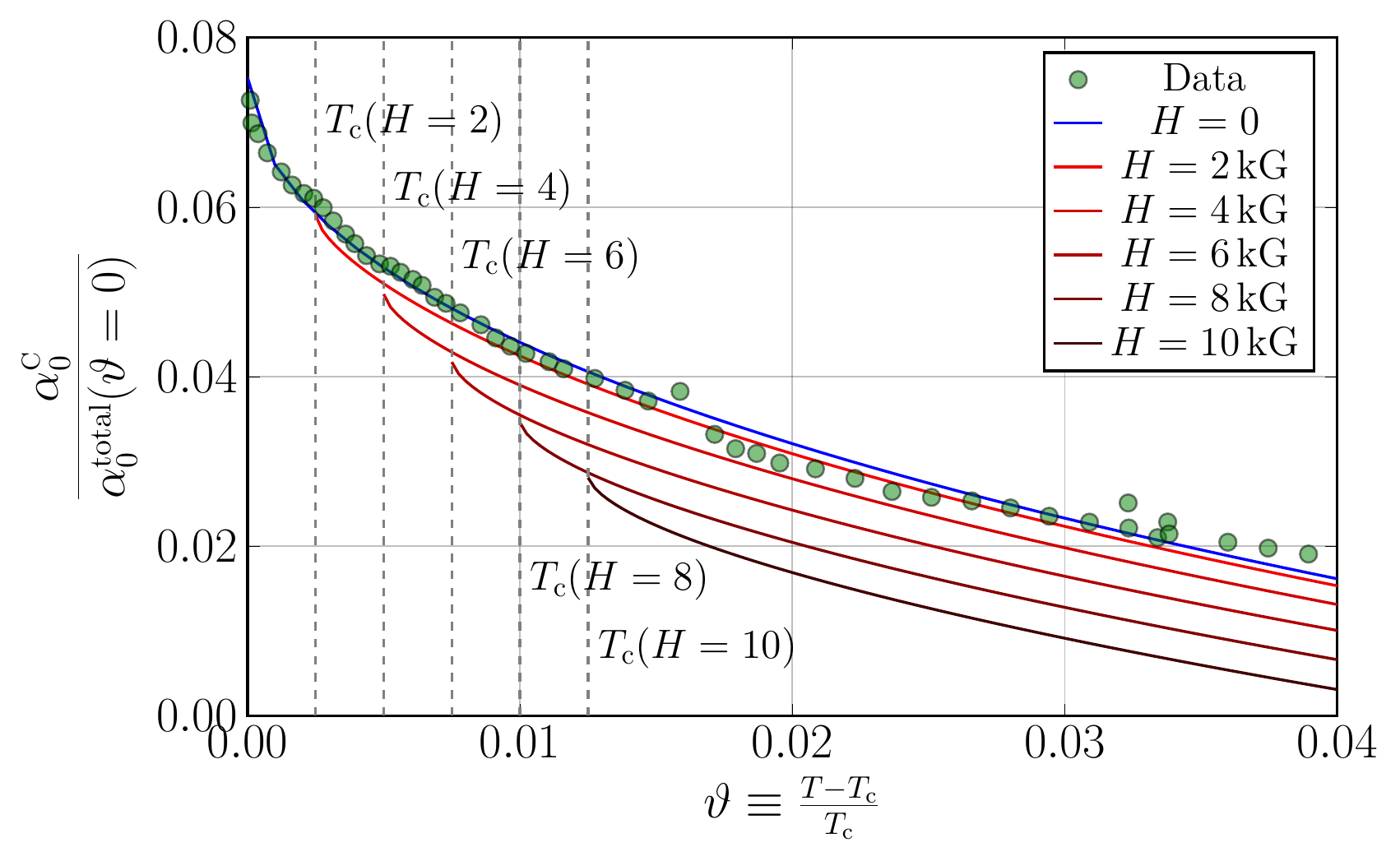}
\caption{
Experimental data (green circles) from Ref.~\onlinecite{pau78c} for the excess attenuation of zero sound at $p=30.87\,\mbox{bar}$ normalized by the total attenuation at $T_c$. 
The theoretical result from Eq.~\eqref{eq-fluctuation_attenuation} shown as the blue curve corresponds to a counter term of $\alpha_{\infty}^{\text{C}}=-0.215\times\alpha_0^{\text{total}}(\vartheta=0)$ and $F^s_2=1.39$, both of which optimize the fit of our theory to the data. 
The solid shades of red curves show the effects of a magnetic field on the quasiparticle-pair- fluctuation contribution to the excess attenuation for spin-triplet pairing.
}
\label{fig-attenuation-theory_vs_experiment}
\end{figure}

For an isotropic Fermi liquid the Landau interaction function can be expressed as a an expansion in Legendre polynomials, $F^s(\hat\vp\cdot\hat\vp')=\sum_{l\ge 0}F^s_l\,P_l(\hat\vp\cdot\hat\vp')$, where $F^s_l$ are the Landau parameters. Similarly, the deformation of the Fermi surface is expressed in terms of spherical harmonics, $\nu_{\hat\vp}=\sum_{lm}\nu_{lm}Y_{lm}(\hat\vp)$. For \He\ the Landau parameters are large and repulsive, $F^s_0\approx 87$, $F^s_1\approx 14$, $F^s_2\approx 1$ and $|F^s_l| < 1$ for $l\ge 3$.~\cite{hal90} This hierarchy implies that zero sound mode is predominantly a dilation of the Fermi surface, $\nu_{00}$, corresponding to density oscillations, $\delta n(\vk,\omega) = 2N(0)\nu_{00}$, coherently phased with the dipole mode, $\nu_{10}$, corresponding to longitudinal current oscillations, $\hat\vk\cdot\vj(\vk,\omega)=2N(0)\nu_{10}$.
The dispersion relation for zero sound, obtained as the eigenfrequency of Eq.~\eqref{eq-collisionless_sound}, is $\omega = c_0 |\vk|$ with a phase velocity that is large compared to the Fermi velocity. In fact $c_0$ is larger than the hydrodynamic sound velocity, $c_1=\nicefrac{1}{\sqrt{3}}v_f[(1+F^s_0)(1+\nicefrac{1}{3}F^s_1)]^{1/2}$, even though quasiparticle collisions no longer contribute to the restoring force for density fluctuations, 
$c_0=c_1\left[1+\nicefrac{4}{5}(1+\nicefrac{1}{5}F^s_2)/(1+F^s_0)\right]^{1/2}$ to leading order in $(v_f/c_0)^2\ll 1$.\cite{bay78} 
Zero sound is weakly damped by thermal quasiparticles scattering off the coherent particle-hole excitation of the Fermi surface, and can by calculated to leading order in $1/\omega\tau_{\text{N}}$ by evaluating the collision integral with the collisionless zero-sound eigenfunction,~\cite{cor69}
\vspace*{-3mm}
\begin{equation}
\hspace*{-2mm}\alpha_0(T) = \frac{2}{15}
              \frac{m^*}{m}
              \left(\frac{v_f}{c_0}\right)^2\ns
                   (1+\nicefrac{1}{5}F^s_2)^2\,\frac{1}{c_0\tau_0}
            = a_0\,T^2
\,,
\vspace*{-3mm}
\end{equation}
where the collision time $\tau_0\propto T^{-2}$ is of proportional to the quasiparticle lifetime $\tau_{\text{N}}(T)$ and a normalized and weighted average of the scattering probability $W_{\text{N}}$ over the Fermi surface.~\cite{bay78}
The result for $\alpha_0(T)$ is well established experimentally, is the signature of the crossover to zero sound propagation at low temperatures,~\cite{abe66} and provides the baseline to study pairing fluctuations via zero sound attenuation.

As $T\rightarrow T_c^+$ particle-hole excitations comprising zero sound interact with long-lived Cooper pair fluctuations leading to an increase in attenuation. 
The pair fluctuation scattering contribution to the collision integral derived from Eq.~\eqref{eq-collision_integral1}, but evaluated with the self energies, $\Sigma_{\vp}^{(d)12}$ and $\Sigma_{\vp}^{(d)21}$, the particle-particle vertex from Eq.~\eqref{eq-Bethe-Salpeter_Cooper_Channel} and the scattering probability from Eqs.~\eqref{eq-Scattering_probability_QP-Pair} and \eqref{eq-Cooper_Pair_propagator}, is given by~\cite{lin21}
\begin{widetext}
\vspace*{-3mm}
\begin{eqnarray}
I_{\vp}^{\text{C}} \ns=\ns 
\int\ns\dthree{p'}
\int\ns\dthree{Q}
W_{\text{C}}(\vp,\vp';\vQ,\eps_{\vp'}+\eps_{-\vp'+\vQ})
\Big[ 
n(\eps_{\vp'})n(\eps_{-\vp'+\vQ})
\left(1-n(\eps_{\vp'}+\eps_{-\vp'+\vQ}-\eps_{\vp})\right)
\left(1-n(\eps_{\vp})\right)
\nonumber\\
-
\left(1-n(\eps_{\vp'})\right)
\left(1-n(\eps_{-\vp'+\vQ})\right)
n(\eps_{\vp'}+\eps_{-\vp'+\vQ}-\eps_{\vp})
n(\eps_{\vp})
\Big]
\delta(\eps_{\vp'}+\eps_{-\vp'+\vQ}-\eps_{\vp}-\eps_{-\vp+\vQ})
\,.\quad
\label{eq-collision_integral-QP_CP}
\end{eqnarray}
\vspace*{-4mm}
\end{widetext}
This result for quasiparticle-pair-fluctuation scattering differs fundamentally from Emery's phenomenological collision integral; $I^{\text{C}}_{\vp}$ is obtained from a Keldysh formulation of transport theory combined with the leading correction to Fermi liquid theory for the self-energy, i.e. the diagram in Fig.~\ref{fmf-Self_Energies}(d). Emery's approach assumes that pair fluctuations with different quantum numbers are phase coherent and that their amplitudes add coherently. 
Our result in Eq.~\eqref{eq-collision_integral-QP_CP} has a smaller prefactor than that of Emery,\footnote{Emery's formula would correspond to a diagram similar to Fig.~\ref{fmf-Self_Energies}(c), but with $\Gamma\rightarrow\Gamma_Q$ \emph{twice}, and the internal particle-hole pair replaced by a particle-particle pair, thus double counting the particle-particle ladder diagrams, which explains why Emery's expression also has a prefactor a factor of 2 larger than our prefactor.} and contains \emph{no quantum interference terms} between pair fluctuations with different quantum numbers because above $T_c$ these fluctuations are not phase coherent.

The pair fluctuation correction to the zero sound attenuation is calculated by evaluating Eq.~\eqref{eq-collision_integral-QP_CP} with the collisionless zero-sound eigenfunction along the lines outlined in Ref.~\onlinecite{cor69}. The correction to the attenuation increases as $T\rightarrow T_c^+$, and is dominated by pair-fluctuation modes with $Q<\pi/\xi_0$ and quasiparticles with energies, $\eps_{\vp}$ near the Fermi energy. In $d=3$ the divergence in $W_{\text{C}}$ is integrable leading to the result,
\begin{eqnarray}
\alpha_0^{\text{C}} 
&=& 
\cA_0\,\Theta(\vartheta) + \alpha_{{\infty}}^{\text{C}}
\,,
\label{eq-fluctuation_attenuation}
\end{eqnarray}
where the prefactor,
\vspace*{-2mm}
\begin{eqnarray}
\hspace*{-5mm}
\cA_0 = 
\frac{2.24}{10\pi}
\frac{1}{N(0)\xi_0^3 T_c}
\Big(\frac{v_f}{c_0}\Big)^3 
\frac{m^*}{m}
\left(1+\nicefrac{1}{5}F_2^s\right)^2
\frac{T_c}{v_f}
\,,
\label{eq-prefactor}
\end{eqnarray}
is of order $10\,\mbox{m}^{-1}$ and increases with pressure, c.f.Fig.~\ref{fig-attenuation-pressure} in the Supplementary Material (SM).\footnote{See Supplemental Material below for 
                details of analysis 
                supporting the corrections to the transport equation, 
                application of the theory to the anomalous attenuation of 
                zero sound for $T\rightarrow T_{c}^+$, and comparison with the
		cutoff procedure of Samalam and Serene.~\cite{sam78}}
The temperature dependence is given by the function 
\vspace*{-2mm}
\begin{equation}
\Theta(\vartheta) 
=
\int_{-\infty}^{\infty}\frac{dx}{\pi}\,\cF(x;\vartheta)\,\sech^4(2x/\pi)
\,,
\label{eq-Theta}
\end{equation}
\vspace*{-3mm}
where 
\vspace*{-3mm}
\begin{equation}
\hspace*{-2mm}
\cF
\equiv
\xi_{1m}^3\ns\int_0^{\infty}\ns\ns Q^2 dQ|\cC_{1m}(Q,\omega)|^2 
\ns=\ns 
\frac{\pi}{2}
\frac{{\sf Im}\,\sqrt{\vartheta + i|x|}}{|x|}
\,, 
\label{eq-mode_integral}
\end{equation}
with $x\equiv\pi\omega/8T$.
The integral defining the temperature dependence, $\Theta(\vartheta)$, is calculated numerically. Note that $\Theta(0)$ is finite and includes contributions to the attenuation at $T_c$ from short wavelengths, $Q\gg \pi/\xi_0$, and thus far from the pole of pair propagator, and the region of validity of the long-wavelength expression for $\cC_{lm}(Q,\omega)$. Thus, we introduce the counter term, $\alpha^{\text{C}}_{\infty}$, to correct for the short wavelength contributions to $\Theta(\vartheta)$ based on Eq.~\eqref{eq-Cooper_Pair_propagator}. The counter term is then determined by the measured value of the excess attenuation at $T_c$. Additional details of this calculation are included in the SM.\cite{Note3}

The excess attenuation of zero sound reported by Paulson and Wheatley~\cite{pau78c} ranges from 2--8\% percent (largest at melting pressure) of the baseline attenuation from Fermi liquid theory, consistent with the $\cO(\sml^3)$ contribution from the quasiparticle-pair-fluctuation self-energy terms in Fig.~\ref{fmf-Self_Energies}(d). 
Note that the pressure dependence of the prefactor $\cA_0$ shown in Fig.~\ref{fmf-Self_Energies} of the SM~\cite{Note3} is consistent with the report by Paulson and Wheatley.~\cite{pau78c}
Furthermore, the temperature dependence of the excess attenuation is in excellent agreement with the prediction of Eqs.~\eqref{eq-fluctuation_attenuation}-\eqref{eq-mode_integral}, as shown in Fig.~\ref{fig-attenuation-theory_vs_experiment}.
In the SM~\cite{Note3} we compare our renormalization procedure with the cutoff procedure used by Samalam and Serene.~\onlinecite{sam78}

%
\noindent{\it Zeeman Splitting --}
A further test of our theory would be to apply a uniform magnetic field to lift the degeneracy for pairing in the three spin-triplet states. The factor $(2S+1)$ in Eq.~\eqref{eq-Cooper_Pair_propagator} is then replaced by a sum over the magnetic quantum number $m_s\in\{+1,0,-1\}$, i.e. $\cC_{lm}(Q,\omega)\rightarrow\cC^{m_s}_{lm}(Q,\omega)$ where the nuclear Zeeman splitting leads to $\vartheta\rightarrow\vartheta_{m_s}= (T-T_{c_{m_s}})/T_{c_{m_s}}$ with $T_{c_{\pm 1}}=T_c\pm\lambda\,H$ and $T_{c_0}=T_c-\eta\,H^2$.
The parameters $\lambda$ and $\eta$ are known; near melting pressure $\lambda\simeq 3.0\mu\mbox{K}/\mbox{kG}$ and $\eta=T_c\,g_z/N(0)\simeq 2.4\,\mu\mbox{K}/\mbox{kG}^2$.\cite{dobbs00}
Thus, the onset superfluidity now corresponds to $T_{c_+}>T_c$, while the instability temperatures for the $m_s=0$ and $m_s=-1$ pairs are suppressed. This results in a reduction of quasiparticle-pair-fluctuation scattering at all temperatures above $T_{c_+}$ as shown for several values of the magnetic field in Fig.~\ref{fig-attenuation-theory_vs_experiment}. Observation of the field dependence of the enhanced attenuation would provide strong support for the theory quasiparticle scattering by long lived pairing fluctuations near $T_c$.

%
\noindent{\it Summary \& Outlook --}
We derived the leading corrections to the Boltzmann-Landau transport equation for an interacting Fermi liquid in the vicinity of the Cooper instability based on Keldysh field theory and the quasiparticle self energies classified to $\cO(\sml^3)$ in the expansion parameters of Fermi liquid theory.
The leading corrections result from the scattering of quasiparticles by long-lived pair fluctuations for temperatures $T\rightarrow T_c^+$.
Application of the theory to calculate the enhanced attenuation of zero sound in liquid \He\ provides strong support for the feedback role of quasiparticle-pair-fluctuation scattering near $T_c$.
The theory is applicable to BCS superconductors and superfluids in the clean limit. An extension to normal Fermi liquids with disorder and Zeeman fields, including liquid \He\ infused into dilute random solids, is for future studies.

\noindent{\it Acknowledgements --} This work was supported by the National Science Foundation Grant DMR-1508730.

%

\newpage

\onecolumngrid

\renewcommand\theequation{SM.\arabic{equation}}
\centerline{\large Supplemental Material}

\bigskip

\subsection{Cooper pair fluctuation contribution to the zero sound attenuation} 

To calculate the zero sound attenuation, we follow the perturbation method used by Corruccini et~al.~\cite{cor69s} In Fourier space the Boltzmann-Landau transport equation, including the collision integral, can be expressed in terms of the deformation of the Fermi surface,
\begin{equation}\label{eq-Boltzmann-Landau_equation-supp}
(\omega - \vv_{\hat\vp}\cdot\vk)\,\nu_{\hat\vp} - (\vv_{\hat\vp}\cdot\vk)
\int\frac{d\Omega_{\hat\vp'}}{4\pi} F^{\mathrm{s}}(\hat\vp\cdot\hat\vp')\,\nu_{\hat\vp'}
= 
i\int d\eps_{\vp}\,I^{\mbox{\tiny total}}_{\vp}[\nu_{\hat\vp'}]
\,,
\end{equation}
where $I^{\mbox{\tiny total}}_{\vp}[\nu_{\hat\vp'}]$ is the collision integral derived from diagrams Fig.~1 (c) and (d) in the Letter, i.e. the binary collision integral defined in Eq.~(4) of the Letter, leading to the normal state transport coefficients of a normal Fermi liquid, and the quasiparticle-pair-fluctuation collision integral defined in Eqs.~( 11), (7) and (8) in the Letter.
The Fermi surface deformation is defined in terms of the deviation of the quasiparticle distribution function, $\nu_{\hat\vp}=\int d\eps_{\vp}\delta n_{\vp}$. Zero sound propagates when the period of oscillation is much shorter than the relaxation time $\tau$. Thus $1/\omega\tau$ is small parameter and we can treat the collision integral as a perturbation to the terms of the collisionless Boltzmann-Landau equation. We expand the distribution function in powers of $1/\omega\tau$, $\nu_{\vp}=\nu^{(0)}_{\vp}+ \nu^{(1)}_{\vp}+\ldots$, 
and similarly for the dispersion relation, $s\equiv \omega/kv_f = s^{(0)} + s^{(1)} + \ldots$, 
\begin{equation}
s\simeq\frac{\omega}{k_0 v_f}\,\left(1- i\,\frac{\alpha_0}{k_0}\right)
\,.
\end{equation}
The attenuation coefficient $\alpha_0$ is defined by the imaginary part of the dispersion relation expressed in terms of the complex wavenumber, $k(\omega)=k_0+i\alpha_0$, where $\lambda=2\pi/k_0$ is the wave length of the sound wave at frequency $\omega$.
Comparing with the perturbation expansion we have $s^{(0)}=\omega/k_0 v_f = c_0/v_f$, and 
\begin{equation}\label{eq-attenuation}
\alpha_0 = i\frac{s^{(1)}}{s^{(0)}}k_0.
\end{equation}
Hereafter we drop the subscript and identify $k_0$ as $k$.
The first order correction to the dispersion relation, $s^{(1)}$, is obtained from Eq.~\eqref{eq-Boltzmann-Landau_equation-supp} and the eigenfunction of the collisionless Boltzmann-Landau transport equation, Eq.~(9) of the Letter,
\begin{equation}\label{eq-s1}
s^{(1)} = \frac{i}{k v_f}\times
\frac
{\displaystyle\int\frac{d\Omega_{\hat\vp}}{4\pi}
\left\{
\int d\eps_{\vp}\,I^{(1)}_{\vp}\left[\psi_{\hat\vp'}^{(0)}\right]
\right\}
\psi_{\hat\vp}^{(0)}
}
{\displaystyle\sum_{l\ge 0}
\frac{(\nu_l^{(0)})^2}{2l+1}\Big(1+\frac{F_l^{s}}{2l+1}\Big)
},
\end{equation} 
where $I^{(1)}_{\vp}$ corresponds to the collision integral \emph{linearized} about the local equilibrium quasiparticle distribution function, $n_0(\eps_{\vp}+\delta\eps_{\vp}(\vk,\omega))$. This is because collision processes are local in space-time and conserve the total quasiparticle energy locally, not the unperturbed quasiparticle energy. The deformation of the Fermi surface expressed in terms of the local deviation of the distribution function is defined as $\psi_{\hat\vp}$, and is related to $\nu_{\hat\vp}$ by
\begin{equation}
\psi_{\hat\vp}=\nu_{\hat\vp} + \int\frac{d\Omega_{\hat\vp'}}{4\pi}
F^{s}(\hat\vp\cdot\hat\vp')\,\nu_{\hat\vp'}
\,.
\end{equation}
Evaluating Eq.~\eqref{eq-s1} with the binary scattering collision integral, Eq.~(4) in the Letter yields the known result for the zero sound attenuation given in Eq.~(10) of the Letter.~\cite{bay78s} The correction from quasiparticle-pair-fluctuation scattering is then obtained by evaluating Eq.~\eqref{eq-s1} with the \emph{linearized form} of collision integral for quasiparticle-pair-fluctuation scattering given by Eqs.~(11),(7) and (8),
\begin{eqnarray}
I^{\text{C}(1)}_{\vp} 
&=&
-\int\frac{d^3 Q}{(2\pi)^3}\int\frac{d^3 p'}{(2\pi)^3}
W_{\text{C}}(\vp,\vp';\vQ,\omega)
\,
\Big(\psi_{\hat\vp}^{(0)}+\psi_{-\hat\vp}^{(0)}-\psi_{\hat\vp'}^{(0)}-\psi_{-\hat\vp'}^{(0)}\Big)
\nonumber\\
&\times& 
\beta
\big[n_0(\eps_{\vp})n_0(\eps_{-\vp})\big(1-n_0(\eps_{\vp'})\big)\big(1-n_0(\eps_{-\vp'})\big)\big]
\delta(\eps_{\vp_{\vp}} + \eps_{-\vp} - \eps_{\vp'} - \eps_{-\vp'})
\,,
\label{eq-linearized_qp-fluc_collision_integral}
\end{eqnarray}
where $W_{\text{C}}(\vp,\vp';\vQ,\omega)$ is the quasiparticle-pair-fluctuation scattering probability given by Eqs.~[7-8] in the Letter. We also set $Q=(\vQ,\omega)\rightarrow (\mathbf{0},0)$, except in the Cooper pair propagator which is singular in the limit $Q\rightarrow 0$. This is justified because these corrections are higher order in $Q^2$ and only suppress the singular contribution from the pair propagator. 
The energy conserving delta function sets $\omega=2\eps_{\vp}$, and since $\eps_{-\vp}=\eps_{\vp}$ 
Eq.~\eqref{eq-linearized_qp-fluc_collision_integral} reduces to
\begin{eqnarray}
\hspace*{-7mm}I^{\text{C}(1)}_{\vp} =
-\frac{\beta}{2}\int\frac{d^3 Q}{(2\pi)^3}N(0)\ns\int\frac{d\Omega_{\vp'}}{4\pi}
W_{\text{C}}(\hat\vp,\hat\vp';\vQ,2\eps_{\vp})
\Big(\psi_{\hat\vp}^{(0)}+\psi_{-\hat\vp}^{(0)}-\psi_{\hat\vp'}^{(0)}-\psi_{-\hat\vp'}^{(0)}\Big)
\left[n_0(\eps_{\vp})\big(1-n_0(\eps_{\vp})\big)\right]^2
\ns,
\label{eq-linearized_qp-fluc_collision_integral2}
\end{eqnarray}
where $\psi_{\hat\vp}=\sum_{l\ge 0}\psi_{l0}\,P_l(\hat\vp\cdot\hat\vk)$; note that $\vk$ is the wavevector of the of the zero sound mode and provides the quantization axis for expansion of the zero sound eigenfunction. Since the zero sound excitation energy is degenerate with respect to the direction of propagation $\hat\vk$ will drop out after averaging over $\hat\vp$, $\hat\vp'$ and $\hat\vQ$. 
The transition probability mediated by pair fluctuations is expanded in eigenfunctions of the Cooper pair fluctuations, and these fluctuations are defined by spherical harmonics with $\vQ$ providing the quantization axis, 

$W_{\text{C}}(\vp,\vp';\vQ$ in terms of the orbital states of the pair fluctuations in Eq. [7] of the Letter.
For zero sound in \He\ the deformation of the Fermi surface is dominated by the spherical harmonics with $l\le 2$; higher harmonics are smaller in magnitude by a factor of order $(v_f/c_0)^2$.~\cite{bay78s}, 
Furthermore, the conservation of particle number and momentum during collisions means that $\psi_{00})$ and $\psi_{10}$ do not contribute to the collision integral, Thus, we can set $\psi_{\pm\hat\vp}=\psi_{20}\,P_(\hat\vp\cdot\hat\vk)$, and $\psi_{\pm\hat\vp'}=\psi_{20}\,P_2(\hat\vp'\cdot\hat\vk)$, in which case Eq.~\eqref{eq-linearized_qp-fluc_collision_integral2} reduces to
\begin{eqnarray}
\hspace*{-7mm}I^{\text{C}(1)}_{\vp} =
-\psi_{20}\,N(0)\beta
\left[n_0(\eps_{\vp})\big(1-n_0(\eps_{\vp})\big)\right]^2
\ns\times\ns
\int\frac{d^3 Q}{(2\pi)^3}\ns\int\frac{d\Omega_{\vp'}}{4\pi}
W_{\text{C}}(\hat\vp,\hat\vp';\vQ,2\eps_{\vp})
\Big(
P_2(\hat\vp\cdot\hat\vk)-P_2(\hat\vp'\cdot\hat\vk)
\Big),
\label{eq-linearized_qp-fluc_collision_integral3}
\end{eqnarray}
where 
$\xi_{1,0}^2 = 3\xi_{1,\pm 1}^2 = \frac{9}{5}\xi_0^2 $, with 
$\xi_0^2 = \frac{7\zeta(3)}{48\pi^2}\frac{v_\mathrm{F}^2}{T_\mathrm{c}^2}$.

Combining Eqs.~\eqref{eq-attenuation},~\eqref{eq-s1} and ~\eqref{eq-linearized_qp-fluc_collision_integral3} gives the following multiple integral for the attenuation,
\begin{eqnarray}
\alpha_0^{\text{C}} &=& \frac{1}{c_0\,D}\,\psi_{20}\,N(0)\,
                        \int d\eps_{\vp}\,\beta\left[n_0(\eps_{\vp})(1-n_0(\eps_{\vp})\right]^2\,
\nonumber\\
&\times&
\frac{1}{2\pi^2}\,\int_0^{\infty}dQ\,Q^2\,\int\dangle{\vQ}\int\dangle{\vp}\int\dangle{\vp'}
\,W_{\text{C}}(\hat\vp,\hat\vp';\hat\vQ)\,\left[P_2(\hat\vp\cdot\hat\vk) - P_2(\hat\vp'\cdot\hat\vk)\right]\,\psi_{\hat\vp}
\,,
\end{eqnarray}
where 
\begin{equation}\label{eq-Denom}
D = \sum_{l\ge 0},\frac{\nu_{l}^2}{2l+1}\,\left(1+F^s_l/2l+1\right)
\,.
\end{equation}

Now, $\psi_{\hat\vp} = \sum_{l\ge 0}\psi_{l,0}\,P_l(\hat\vp\cdot\hat\vk)$ for longitudinal zero sound with wave vector $\vk=k\hat\vk$. Furthermore for zero sound in \He\ only the terms with $l\le 2$ are relevant; terms with $l\ge 3$ are smaller in magnitude by at least $(v_f/c_0)^2$. Thus, $\psi_{\hat\vp} = \psi_{0,0} + \psi_{1,0}\,(\hat\vp\cdot\hat\vk) + \psi_{2,0}\,P_2(\hat\vp\cdot\hat\vk) + \mbox{negligible}$.
The conservation laws for particle number and momentum for collision events ensure that the terms $\psi_{0,0}$ and $\psi_{1,0}$ drop out; $\psi_{0,0}$ drops out because $W_{\text{C}}(\hat\vp,\hat\vp';\hat\vQ)=W_{\text{C}}(\hat\vp',\hat\vp;\hat\vQ)$, and $\psi_{1,0}$ drops out because $W_{\text{C}}(-\hat\vp,\hat\vp';\hat\vQ)=W_{\text{C}}(\hat\vp',\hat\vp;\hat\vQ)$, and $P_2(-\hat\vp\cdot\hat\vk)=P_2(\hat\vp\cdot\hat\vk)$.
Thus, we obtain
\begin{eqnarray}\label{eq-alpha0C-double_integral}
\alpha_0^{\text{C}} &=& \frac{\psi_{2,0}^2}{c_0\,D}\,N(0)\,
\int d\eps_{\vp}\,\beta\left[n_0(\eps_{\vp})(1-n_0(\eps_{\vp})\right]^2\,
\times\frac{1}{2\pi^2}\,\int_0^{\infty}dQ\,Q^2\,
\nonumber\\
&\times&
\int\dangle{\vp}\int\dangle{\vp'}\int\dangle{\vQ}
\,W_{\text{C}}(\hat\vp,\hat\vp';\hat\vQ)\,
\left[P_2(\hat\vp\cdot\hat\vk) - P_2(\hat\vp'\cdot\hat\vk)\right]\,P_2(\hat\vp\cdot\hat\vk)
\,.
\end{eqnarray}
It is most efficient to carry out the angular integration over the pair fluctuation wavevector first, since $\hat\vQ$ appears only in the transition probability for quasiparticle-pair-fluctuation scattering. In particular we can express
\begin{equation}
W_{\text{C}} = \frac{27\pi}{N(0)^2}\,
\left\{
\nicefrac{1}{2}|\cC_{1,1}|^2
-
\nicefrac{1}{2}|\cC_{1,1}|^2\,\left[(\hat\vp\cdot\hat\vQ)^2+(\hat\vp'\cdot\hat\vQ)^2\right]
+
\Big(|\cC_{1,0}|^2 + \nicefrac{1}{2}|\cC_{1.1}|^2\Big)\,(\hat\vp\cdot\hat\vQ)^2(\hat\vp'\cdot\hat\vQ)^2
\right\}
\,,
\end{equation} 
and carry out the integration over $\Omega_{\vQ}$ using 
\begin{equation}
\int\dangle{\vQ}\hat\vQ_i\hat\vQ_j = \frac{1}{3}\,\delta_{ij}
\,\,\mbox{and}\,\, 
\int\dangle{\vQ}
\hat\vQ_i\hat\vQ_j\hat\vQ_k\hat\vQ_l = \frac{1}{15}\,
\left(\delta_{ij}\delta_{kl} + \delta_{ik}\delta_{jl} + \delta_{il}\delta{jl}\right)
\,.
\end{equation}
This yields, 
\begin{equation}
\langle W_{\text{C}}\rangle_{\hat\vQ} \equiv \int\dangle{\vQ}\,W_{\text{C}} 
= 
\frac{27\pi}{N(0)^2}\,
\left\{
\nicefrac{1}{15}|\cC_{1,0}|^2 + \nicefrac{1}{5}|\cC_{1,1}|^2 +
\nicefrac{2}{15}|\left(|\cC_{1,0}|^2 + \nicefrac{1}{2}|\cC_{1,1}|^2\right)\,(\hat\vp\cdot\hat\vp')^2
\right\}
\,.
\end{equation}
The remaining angular integrations of the intermediate and scattered quasiparticle momentum reduce to  
\begin{equation}
\bar\omega_{\text{C}} 
\equiv 
\langle\langle
\langle W_{\text{C}}\rangle_{\hat\vQ}
\left[P_2(\hat\vp\cdot\hat\vk) - P_2(\hat\vp'\cdot\hat\vk)\right]\,P_2(\hat\vp\cdot\hat\vk)
\rangle_{\hat\vp'}\rangle_{\hat\vp}
\,.
\end{equation}
These angular integrations are easily carried out using the identities, 
\begin{equation}
\langle P_2(\hat\vp\cdot\hat\vk)^2\rangle_{\hat\vp}=\nicefrac{1}{15}
\quad \mbox{and}\quad 
\langle P_2(\hat\vp\cdot\hat\vk)\,\hat\vp_i\hat\vp_j\rangle_{\hat\vp}=
-\nicefrac{1}{15}\delta_{ij} + \nicefrac{1}{5}\hat\vk_i\hat\vk_j
\,.
\end{equation}
The angular averaged transition probability then reduces to
\begin{equation}
\bar\omega_{\text{C}} = \frac{27\pi}{N(0)^2}\,
\frac{1}{75}
\Big\{
\frac{7}{5}|\cC_{1,0}|^2 
+ 
2\times
\frac{8}{5}
|\cC_{1.1}|^2
\Big\}
\,.
\end{equation}

\begin{figure}[tbp]
\includegraphics[width=0.75\textwidth]{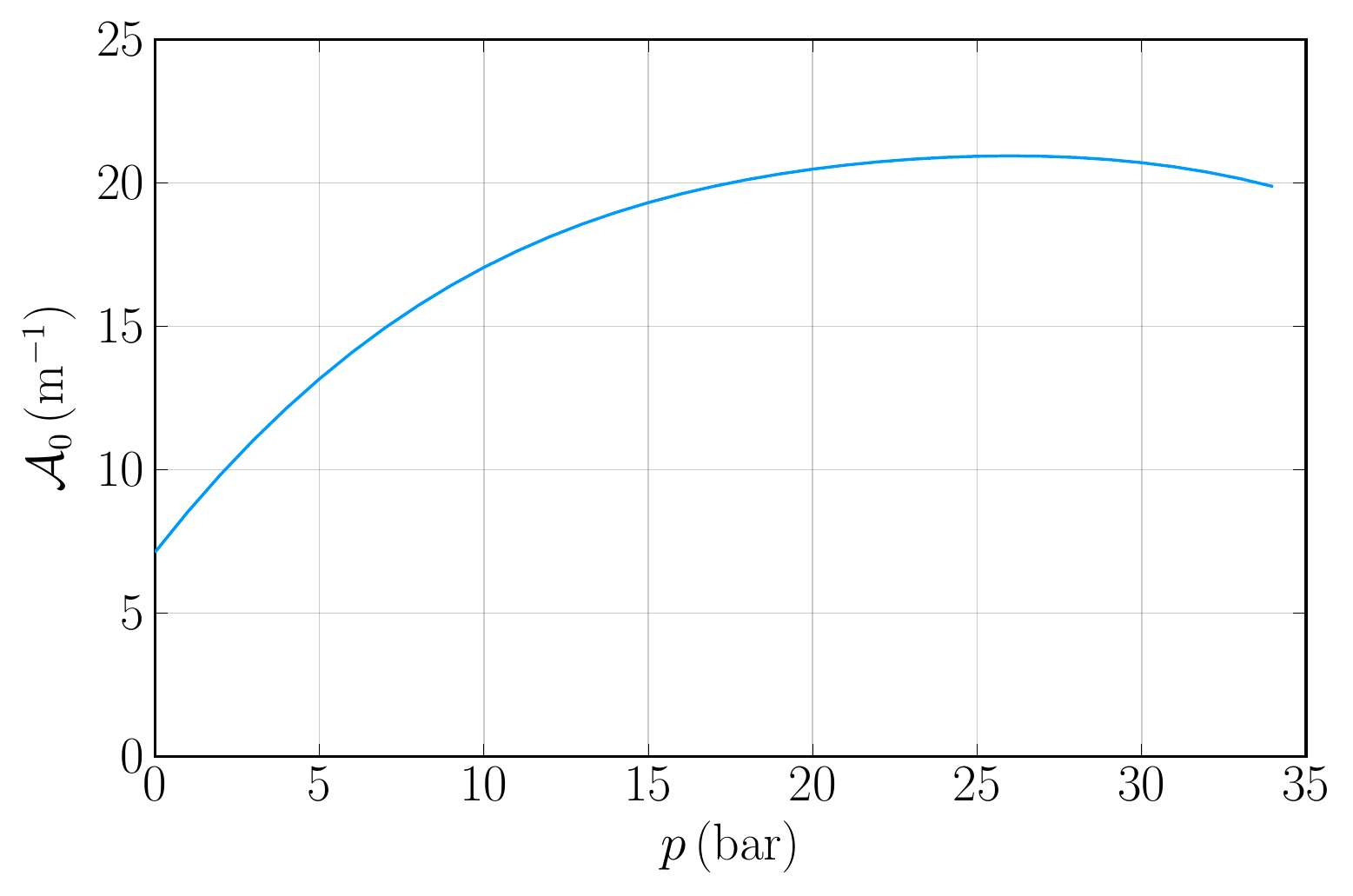}
\caption{
Pressure dependence of the magnitude of the excess attenuation from quasiparticle-pair-fluctuation scattering given by Eq.~(13) in the Letter. The excess attenuation increases by approximately a factor of $4$ from low to high pressure in agreement with the observations of Ref.~[\onlinecite{pau78s}].
} 
\label{fig-attenuation-pressure}
\end{figure}

We can now express the excess attenuation form quasiparticle-pair-fluctuation scattering as an integral
over the pair excitation energy and pair momentum,
\begin{equation}\label{eq-alpha0C-1}
\alpha_0^{\text{C}} 
= 
\frac{\psi_{2,0}^2}{c_0\,D}\,\frac{1}{2\pi^2\,T_c}\,
\Bigg\{
\frac{1}{16}\int_{-\infty}^{+\infty} d\eps_{\vp}\,\sech^4(\eps_{\vp}/2T_c)
\times
\frac{27\pi}{25\,N(0)}\,
\int_0^{\infty}dQ\,Q^2\,
\,
\left[\frac{7}{15}|\cC_{1,0}|^2 + \frac{16}{15} |\cC_{1.1}|^2\right]
\Bigg\}
\,.
\end{equation}

Carrying out the integration over the pair-fluctuation modes first gives,
\begin{eqnarray}
\int_0^{\infty} Q^2 dQ\,|\cC_{1,m}(Q,2\eps_{\vp}))^2 
&=& 
\frac{1}{\xi_{1,m}^3}\,\displaystyle{\cF\left(\frac{\pi\eps_{\vp}}{4T_c};\vartheta\right)}
\,,
\\
\cF(x;\vartheta)
\equiv
\int_0^{\infty}dy\frac{y^2}{(\vartheta + y^2) + x^2}
&=&
\frac{\pi}{2}\frac{\mbox{Im}\sqrt{\vartheta+i|x|}}{|x|}
\,.
\end{eqnarray}
Changing integration variables in Eq.~\eqref{eq-alpha0C-1} to $\eps_{\vp}\rightarrow x=\pi\eps_{\vp}/4T_c$, and recalling that 
$\xi_{1,0}^2 = 3\xi_{1,\pm 1}^2 = \nicefrac{9}{5}\xi_0^2$, yields
\begin{equation}\label{eq-alpha0C-2}
\alpha_0^{\text{C}} 
= 
\frac{N_{\#}}{10\pi}\,
\frac{T_c}{N(0)\xi_0^3\,T_c}\,
\times
\frac{3}{2}\,
\frac{\psi_{2,0}^2}{c_0\,D}\,
\times
\Theta(\vartheta)
\,,
\end{equation}
where
\begin{equation}
\Theta(\vartheta)\equiv \int_0^{\infty}\frac{dx}{\pi}\cF(x;\vartheta)
\,,
\end{equation}
and 
\begin{equation}
N_{\#}\equiv
\frac{9}{2}
\left\{
\frac{7}{75}
\left(\frac{5}{9}\right)^{\nicefrac{3}{2}}
+
2\frac{8}{75}\left(\frac{5}{3}\right)^{\nicefrac{3}{2}}
\right\}
\approx 2.24
\,.
\end{equation}

The denominator from Eq.~\eqref{eq-Denom} is evaluated in terms of the eigenfunction for the collisionless zero sound mode. For \He\ only the terms with $l\le 2$ are substantial. In particular, the hierarchy of interactions: $F^s_0\sim\cO(100)$, $F^s_1\sim \cO(10)$ and $F^s_2\sim\cO(1)$ implies $\nu_{2,0}=2\nu_{0,0}$ and $\nu_{1,0}\simeq 3(c_0/v_f)(m^*/m)\,\nu_{0,0}$ (c.f. p.53 of Baym and Pethick~\cite{bay78s}) Thus, only the $l=0$ and $l=1$ terms contribute significantly. Also using $(c_0/v_f)^2\simeq (c_1/v_f)^2=\nicefrac{1}{3}(1+F^s_0)(m^*/m)$  we obtain,
\begin{equation}
D \simeq \nu_{0,0}^2\,\left(1+F^s_0\right) +
\frac{1}{3}\left(3\frac{c_0}{v_f}\frac{m}{m^*}\nu_{0,0}\right)^2\,\frac{m^*}{m}
\simeq 
2\nu_{0,0}^2\,\left(1+F^s_0\right) \simeq 6\left(\frac{c_0}{v_f}\right)^2\left(\frac{m}{m^*}\right)\,\nu_{0,0}^2 \,.
\end{equation}
Similarly, for the numerator we have $\psi_{2,0}^2=4\,\nu_{0,0}^2\,(1+\nicefrac{1}{5}F^s_2)^2$. Collecting these results, inserting into Eq.~\eqref{eq-alpha0C-2}, and adding the counter term as described in the Letter, we obtain
\begin{eqnarray}\label{eq-alpha0C-3}
\alpha_0^{\text{C}} 
&=& 
\cA_0
\times\Theta(\vartheta)
+
\alpha_{\infty}^{\text{C}}
\,,
\\
\cA_0
&=& 
\frac{N_{\#}}{10\pi}\,
\frac{1}{N(0)\xi_0^3\,T_c}\,
\left(\frac{v_f}{c_0}\right)^3\,
\frac{m^*}{m}\,
\frac{T_c}{v_f}\,
\left(1+\nicefrac{1}{5}F^s_2\right)^2
\,,
\end{eqnarray}
which are equivalent to Eqs.~(12)-(15) in the Letter.

The magnitude and pressure dependence of the excess attenuation is determined by material parameters that enter the coefficient $\cA_0$. Except for $F_2^s$ these material parameters are well known over the full pressure range of liquid \He.\cite{har00s} Using the material parameters obtained from Ref.~\cite{har00s} and the value of $F_2^s=1.39$ obtained by optimizing the fit of the temperature dependence of Eq.~\eqref{eq-alpha0C-3} to the data of Paulson and Wheatley~\cite{pau78s} we obtain the pressure dependence of $\cA_0$ shown in Fig.~\ref{fig-attenuation-pressure}. This result shows that excess attenuation increases by approximately a factor of $4$ over the full pressure range, in good agreement with with that reported by Paulson and Wheatley.~\cite{pau78s}

\subsection{Cutoff Procedure of Samalam and Serene}\label{sec-zero-sound-cutoff}

Samalam and Serene~\cite{sam78s} calculated the attenuation of zero sound in \He\ based on Emergy's heuristic collision integral, motivated by Paulson and Wheatley's report.\cite{pau78s} Their expression for the excess attenuation uses a cutoff procedure to exclude pair-fluctuations with wavevectors $Q\gtrsim \xi_0^{-1}$, with the advantage that it leads to an approximate analytic formula for the excess attenuation that depends on $\sqrt{\vartheta}$ and the cutoff.

Here we apply the procedure of Samalam and Serene to evaluate the double integral in Eq.~\eqref{eq-alpha0C-1} for our result for the transition probability for scattering quasiparticles from pair fluctuations, and compare this approximate result with our renormalization procedure. 

The starting point is the double integral over mode energy and momentum in Eq.~\eqref{eq-alpha0C-double_integral},
\begin{equation}\label{double_integral}
J = \int_{-\infty}^{+\infty} d\eps_{\vp}\,\left[n_0(\eps_{\vp})(1-n_0(\eps_{\vp})\right]^2
    \int_0^{\infty}Q^2 dQ\,\left|\mathcal{C}_{1,m}(Q,2\eps_{\vp})\right|^2
\,,
\end{equation}
Following Samalam and Serene, we approximate the Fermi distribution functions in Eq.~\eqref{double_integral} by their values at $\eps_{\vp}=0$, i.e. $n_0(0)=(1-n_0(0)) = \nicefrac{1}{2}$, then carry out the integration over $\eps_{\vp}$ by closing the contour in the upper half plane to obtain,
\begin{equation}\label{approx}
\int_{-\infty}^{+\infty}d\eps_{\vp} |\mathcal{C}_{1,m}(Q,2\eps_{\vp})|^2 = \frac{4T}{\vartheta + \xi_m^2 Q^2}
\,.
\end{equation}
The remaining phase space integration over the pair-fluctuation modes is now ultraviolet divergent. Samalam and Serence introduce an ultraviolet cutoff to eliminate modes with $\xi_m Q \ge x_c$, with the presumption that $x_c\sim\mathcal{O}(1)$. The resulting integral over modes gives,
\begin{equation}\label{q-integral}
\int_0^{Q_c} dQ\frac{Q^2}{\vartheta + \xi_{1m}^2 Q^2}
= 
\frac{1}{\xi_{1m}^3} \int_0^{x_\mathrm{c}} dx \frac{x^2}{\vartheta + x^2} \\
= 
\frac{1}{\xi_{1m}^3} \Big[x_\mathrm{c} - \sqrt{\vartheta}\arctan(x_\mathrm{c}/\sqrt{\vartheta}) \Big]
\,.
\end{equation}
The angular integrations for the excess attenuation are the same as described in the previous section yielding the the following analytic approximation for the pair-fluctuation contribution to the zero sound attenuation,
\begin{equation}\label{attenuation-cutoff}
\alpha_0^{\mbox{\tiny C}} 
\simeq 
\mathcal{A}_0\,\times\,
\Big[x_\mathrm{c}-\sqrt{\vartheta}\arctan(x_\mathrm{c}/\sqrt{\vartheta})\Big]
\,.
\end{equation}
Note that there is no counter term since the short-wavelength modes have been excluded by the cutoff. 

\begin{figure}[tbp]
\centerline{\includegraphics[width=1.0\textwidth]{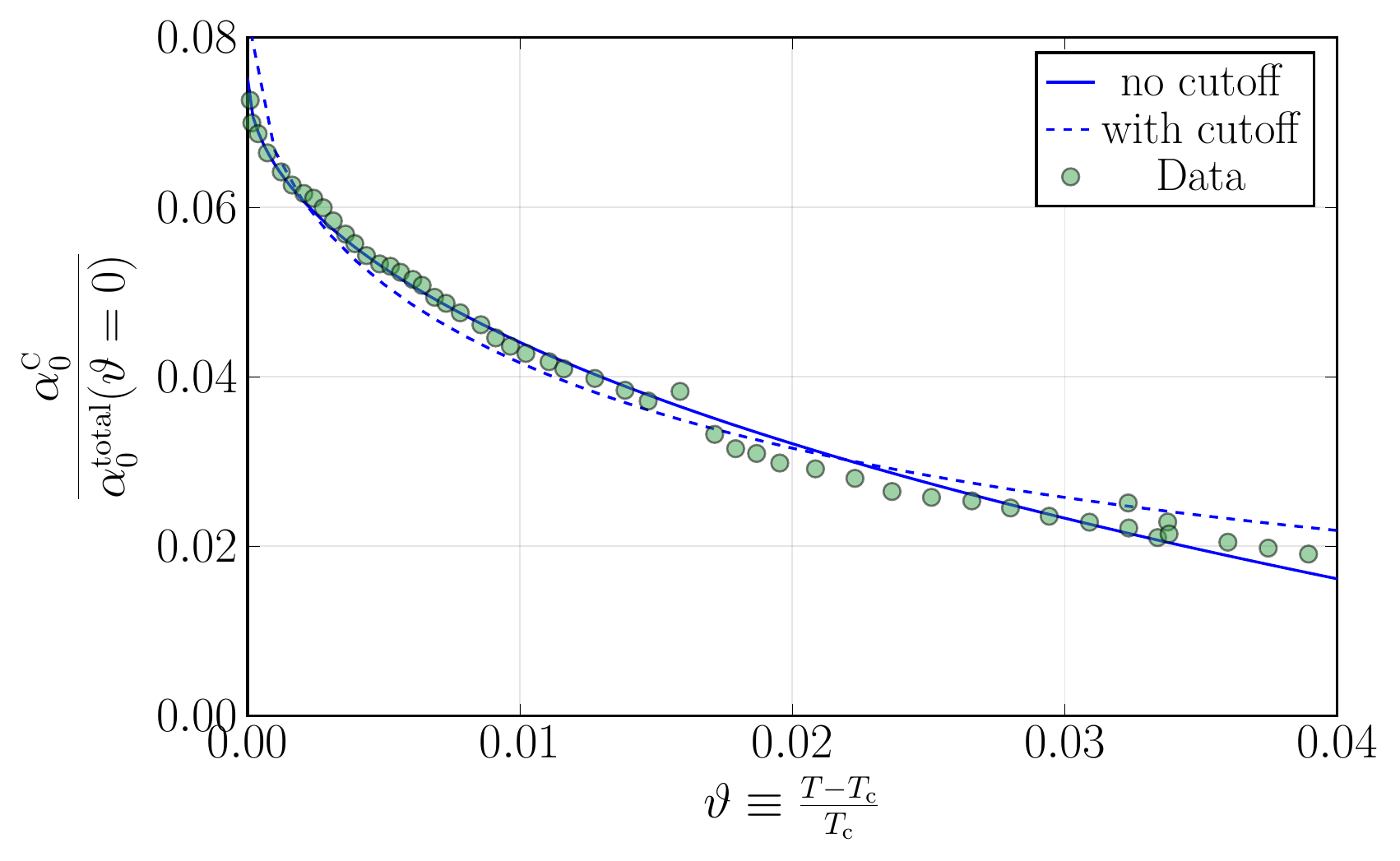}}
\caption{Excess attenuation normalized to the total attenuation at $T_c$. The data are from Ref.~\cite{pau78s}. The solid blue curve is our theoretical prediction from Eq.~\eqref{eq-alpha0C-3}, with the $\alpha_{\infty}^{\text{C}}$ and $F^s_2$ determined by the best fit of the excess attenuation $\alpha^{\mbox{\tiny C}}_{\infty} = -0.215\times\alpha_0^{\text{total}}(\vartheta=0)$ and $F^s_2=1.39$. The dashed blue curve is the best fit using the approximate formula in Eq.~\eqref{attenuation-cutoff} with $x_{\mathrm{c}}=0.236$ and $F^{s}_2=3.25$.}
\label{fig-attenuation-fit}
\end{figure}

Our result based on the cutoff procedure of Samalam and Serene is shown in Fig.~\ref{fig-attenuation-fit} in comparison with our renormalization procedure. The cutoff procedure of Samalam and Serene provides a reasonable representation of the experimental data reported by Paulson and Wheatly~\cite{pau78s}, with a square fit yielding $x_{\mathrm{c}}\approx 0.236$ and $F^{s}_2\approx 3.25$. However, our renormalization procedure gives a near perfect fit to the experimental data with $\alpha_{\infty}^{\mbox{\tiny C}}\approx -0.215\times\alpha_0^{\text{total}}(\vartheta=0)$ and $F^{s}_2\approx 1.39$. We do not have precise data for the value of $F^{s}_2$, but the smaller value of $F^{s}_2\approx 1.39$ given by our renormalization procedure is close to currently reported experimental results.\cite{har00s,dobbs00s}
%
Note that this is not a comparison of our theory directly with the results of Samalam and Serene, which was based Emery's incorrect collision integral, but rather our theoretical result for $\alpha^{\text{C}}_0$, but evaluated using the approximation and cutoff procedure proposed in Ref.~\onlinecite{sam78s}.

\begin{thebibliography}{41}%
\makeatletter
\providecommand \@ifxundefined [1]{%
 \@ifx{#1\undefined}
}%
\providecommand \@ifnum [1]{%
 \ifnum #1\expandafter \@firstoftwo
 \else \expandafter \@secondoftwo
 \fi
}%
\providecommand \@ifx [1]{%
 \ifx #1\expandafter \@firstoftwo
 \else \expandafter \@secondoftwo
 \fi
}%
\providecommand \natexlab [1]{#1}%
\providecommand \enquote  [1]{``#1''}%
\providecommand \bibnamefont  [1]{#1}%
\providecommand \bibfnamefont [1]{#1}%
\providecommand \citenamefont [1]{#1}%
\providecommand \href@noop [0]{\@secondoftwo}%
\providecommand \href [0]{\begingroup \@sanitize@url \@href}%
\providecommand \@href[1]{\@@startlink{#1}\@@href}%
\providecommand \@@href[1]{\endgroup#1\@@endlink}%
\providecommand \@sanitize@url [0]{\catcode `\\12\catcode `\$12\catcode
  `\&12\catcode `\#12\catcode `\^12\catcode `\_12\catcode `\%12\relax}%
\providecommand \@@startlink[1]{}%
\providecommand \@@endlink[0]{}%
\providecommand \url  [0]{\begingroup\@sanitize@url \@url }%
\providecommand \@url [1]{\endgroup\@href {#1}{\urlprefix }}%
\providecommand \urlprefix  [0]{URL }%
\providecommand \Eprint [0]{\href }%
\providecommand \doibase [0]{http://dx.doi.org/}%
\providecommand \selectlanguage [0]{\@gobble}%
\providecommand \bibinfo  [0]{\@secondoftwo}%
\providecommand \bibfield  [0]{\@secondoftwo}%
\providecommand \translation [1]{[#1]}%
\providecommand \BibitemOpen [0]{}%
\providecommand \bibitemStop [0]{}%
\providecommand \bibitemNoStop [0]{.\EOS\space}%
\providecommand \EOS [0]{\spacefactor3000\relax}%
\providecommand \BibitemShut  [1]{\csname bibitem#1\endcsname}%
\let\auto@bib@innerbib\@empty
\bibitem [{\citenamefont {Glover}(1967)}]{glo67}%
  \BibitemOpen
  \bibfield  {author} {\bibinfo {author} {\bibfnamefont {R.}~\bibnamefont
  {Glover}},\ }\bibfield  {{Ideal resistive transition of a superconductor}}
  {\emph {\bibinfo {title} {{Ideal resistive transition of a superconductor}},\
  }}\href {\doibase https://doi.org/10.1016/0375-9601(67)90036-9} {\bibfield
  {journal} {\bibinfo  {journal} {Phys. Lett. A}\ }\textbf {\bibinfo {volume}
  {25}},\ \bibinfo {pages} {542} (\bibinfo {year} {1967})}\BibitemShut
  {NoStop}%
\bibitem [{\citenamefont {Skocpol}\ and\ \citenamefont
  {Tinkham}(1975)}]{sko75}%
  \BibitemOpen
  \bibfield  {author} {\bibinfo {author} {\bibfnamefont {W.~J.}\ \bibnamefont
  {Skocpol}}\ and\ \bibinfo {author} {\bibfnamefont {M.}~\bibnamefont
  {Tinkham}},\ }\bibfield  {{Fluctuations near superconducting phase
  transitions}} {\emph {\bibinfo {title} {{Fluctuations near superconducting
  phase transitions}},\ }}\href {\doibase 10.1088/0034-4885/38/9/001}
  {\bibfield  {journal} {\bibinfo  {journal} {Reports on Progress in Physics}\
  }\textbf {\bibinfo {volume} {38}},\ \bibinfo {pages} {1049} (\bibinfo {year}
  {1975})}\BibitemShut {NoStop}%
\bibitem [{\citenamefont {Aslamazov}\ and\ \citenamefont
  {Larkin}(1968)}]{asl68a}%
  \BibitemOpen
  \bibfield  {author} {\bibinfo {author} {\bibfnamefont {L.~G.}\ \bibnamefont
  {Aslamazov}}\ and\ \bibinfo {author} {\bibfnamefont {A.~I.}\ \bibnamefont
  {Larkin}},\ }\bibfield  {{Effect of Fluctuations on the Properties of a
  Superconductor above the Critical Temperature}} {\emph {\bibinfo {title}
  {{Effect of Fluctuations on the Properties of a Superconductor above the
  Critical Temperature}},\ }}\href@noop {} {\bibfield  {journal} {\bibinfo
  {journal} {Sov. Phys. -- Solid State}\ }\textbf {\bibinfo {volume} {10}},\
  \bibinfo {pages} {875} (\bibinfo {year} {1968})}\BibitemShut {NoStop}%
\bibitem [{\citenamefont {Maki}(1968)}]{mak68}%
  \BibitemOpen
  \bibfield  {author} {\bibinfo {author} {\bibfnamefont {K.}~\bibnamefont
  {Maki}},\ }\bibfield  {The {{Critical Fluctuation}} of the {{Order
  Parameter}} in {{Type}}-{{II Superconductors}}} {\emph {\bibinfo {title} {The
  {{Critical Fluctuation}} of the {{Order Parameter}} in {{Type}}-{{II
  Superconductors}}},\ }}\href {\doibase 10.1143/PTP.39.897} {\bibfield
  {journal} {\bibinfo  {journal} {Prog. Theor. Phys.}\ }\textbf {\bibinfo
  {volume} {39}},\ \bibinfo {pages} {897} (\bibinfo {year} {1968})}\BibitemShut
  {NoStop}%
\bibitem [{\citenamefont {Thompson}(1970)}]{tho70}%
  \BibitemOpen
  \bibfield  {author} {\bibinfo {author} {\bibfnamefont {R.~S.}\ \bibnamefont
  {Thompson}},\ }\bibfield  {{Microwave, Flux Flow, and Fluctuation Resistance
  of Dirty Type-II Superconductors}} {\emph {\bibinfo {title} {{Microwave, Flux
  Flow, and Fluctuation Resistance of Dirty Type-II Superconductors}},\ }}\href
  {\doibase 10.1103/PhysRevB.1.327} {\bibfield  {journal} {\bibinfo  {journal}
  {Phys. Rev. B}\ }\textbf {\bibinfo {volume} {1}},\ \bibinfo {pages} {327}
  (\bibinfo {year} {1970})}\BibitemShut {NoStop}%
\bibitem [{\citenamefont {Enss}\ \emph {et~al.}(2011)\citenamefont {Enss},
  \citenamefont {Haussmann},\ and\ \citenamefont {Zwerger}}]{ens11}%
  \BibitemOpen
  \bibfield  {author} {\bibinfo {author} {\bibfnamefont {T.}~\bibnamefont
  {Enss}}, \bibinfo {author} {\bibfnamefont {R.}~\bibnamefont {Haussmann}}, \
  and\ \bibinfo {author} {\bibfnamefont {W.}~\bibnamefont {Zwerger}},\
  }\bibfield  {{Viscosity and Scale Invariance in the Unitary Fermi Gas}}
  {\emph {\bibinfo {title} {{Viscosity and Scale Invariance in the Unitary
  Fermi Gas}},\ }}\href {\doibase 10.1016/j.aop.2010.10.002} {\bibfield
  {journal} {\bibinfo  {journal} {Ann. Phys.}\ }\textbf {\bibinfo {volume}
  {326}},\ \bibinfo {pages} {770} (\bibinfo {year} {2011})}\BibitemShut
  {NoStop}%
\bibitem [{\citenamefont {Enss}\ and\ \citenamefont {Haussmann}(2012)}]{ens12}%
  \BibitemOpen
  \bibfield  {author} {\bibinfo {author} {\bibfnamefont {T.}~\bibnamefont
  {Enss}}\ and\ \bibinfo {author} {\bibfnamefont {R.}~\bibnamefont
  {Haussmann}},\ }\bibfield  {{Quantum Mechanical Limitations to Spin Diffusion
  in the Unitary Fermi Gas}} {\emph {\bibinfo {title} {{Quantum Mechanical
  Limitations to Spin Diffusion in the Unitary Fermi Gas}},\ }}\href {\doibase
  10.1103/PhysRevLett.109.195303} {\bibfield  {journal} {\bibinfo  {journal}
  {Phys. Rev. Lett.}\ }\textbf {\bibinfo {volume} {109}},\ \bibinfo {pages}
  {195303} (\bibinfo {year} {2012})}\BibitemShut {NoStop}%
\bibitem [{\citenamefont {Liu}\ \emph {et~al.}(2014)\citenamefont {Liu},
  \citenamefont {Zhai},\ and\ \citenamefont {Zhang}}]{liu14}%
  \BibitemOpen
  \bibfield  {author} {\bibinfo {author} {\bibfnamefont {B.}~\bibnamefont
  {Liu}}, \bibinfo {author} {\bibfnamefont {H.}~\bibnamefont {Zhai}}, \ and\
  \bibinfo {author} {\bibfnamefont {S.}~\bibnamefont {Zhang}},\ }\bibfield
  {{Fluctuation Effects on the Transport Properties of Unitary Fermi Gases}}
  {\emph {\bibinfo {title} {{Fluctuation Effects on the Transport Properties of
  Unitary Fermi Gases}},\ }}\href {\doibase 10.1103/PhysRevA.90.051602}
  {\bibfield  {journal} {\bibinfo  {journal} {Phys. Rev. A}\ }\textbf {\bibinfo
  {volume} {90}},\ \bibinfo {pages} {051602} (\bibinfo {year}
  {2014})}\BibitemShut {NoStop}%
\bibitem [{\citenamefont {Eschrig}\ \emph {et~al.}(1999)\citenamefont
  {Eschrig}, \citenamefont {Rainer},\ and\ \citenamefont {Sauls}}]{esc99a}%
  \BibitemOpen
  \bibfield  {author} {\bibinfo {author} {\bibfnamefont {M.}~\bibnamefont
  {Eschrig}}, \bibinfo {author} {\bibfnamefont {D.}~\bibnamefont {Rainer}}, \
  and\ \bibinfo {author} {\bibfnamefont {J.~A.}\ \bibnamefont {Sauls}},\
  }\bibfield  {{Effects of Strong Magnetic Fields on the Pairing Fluctuations
  in High Temperature Superconductors}} {\emph {\bibinfo {title} {{Effects of
  Strong Magnetic Fields on the Pairing Fluctuations in High Temperature
  Superconductors}},\ }}\href {\doibase 10.1103/PhysRevB.59.12095} {\bibfield
  {journal} {\bibinfo  {journal} {Phys. Rev. B}\ }\textbf {\bibinfo {volume}
  {59}},\ \bibinfo {pages} {12095} (\bibinfo {year} {1999})}\BibitemShut
  {NoStop}%
\bibitem [{\citenamefont {Chen}\ \emph {et~al.}(2005)\citenamefont {Chen},
  \citenamefont {Stajic}, \citenamefont {Tan},\ and\ \citenamefont
  {Levin}}]{che05a}%
  \BibitemOpen
  \bibfield  {author} {\bibinfo {author} {\bibfnamefont {Q.}~\bibnamefont
  {Chen}}, \bibinfo {author} {\bibfnamefont {J.}~\bibnamefont {Stajic}},
  \bibinfo {author} {\bibfnamefont {S.}~\bibnamefont {Tan}}, \ and\ \bibinfo
  {author} {\bibfnamefont {K.}~\bibnamefont {Levin}},\ }\bibfield  {{BCS-BEC
  Crossover: From High Temperature Superconductors to Ultracold Superfluids}}
  {\emph {\bibinfo {title} {{BCS-BEC Crossover: From High Temperature
  Superconductors to Ultracold Superfluids}},\ }}\href {\doibase
  10.1016/j.physrep.2005.02.005} {\bibfield  {journal} {\bibinfo  {journal}
  {Phys. Rep.}\ }\textbf {\bibinfo {volume} {412}},\ \bibinfo {pages} {1}
  (\bibinfo {year} {2005})}\BibitemShut {NoStop}%
\bibitem [{\citenamefont {Landau}(1957)}]{lan57}%
  \BibitemOpen
  \bibfield  {author} {\bibinfo {author} {\bibfnamefont {L.~D.}\ \bibnamefont
  {Landau}},\ }\bibfield  {{Oscillations in a Fermi Liquid}} {\emph {\bibinfo
  {title} {{Oscillations in a Fermi Liquid}},\ }}\href@noop {} {\bibfield
  {journal} {\bibinfo  {journal} {Sov. Phys. JETP}\ }\textbf {\bibinfo {volume}
  {32}},\ \bibinfo {pages} {59} (\bibinfo {year} {1957})}\BibitemShut {NoStop}%
\bibitem [{\citenamefont {Abrikosov}\ and\ \citenamefont
  {Khalatnikov}(1957)}]{abr57}%
  \BibitemOpen
  \bibfield  {author} {\bibinfo {author} {\bibfnamefont {A.~A.}\ \bibnamefont
  {Abrikosov}}\ and\ \bibinfo {author} {\bibfnamefont {I.}~\bibnamefont
  {Khalatnikov}},\ }\bibfield  {{Theory of Kinetic Phenomena in Liquid $^3$He}}
  {\emph {\bibinfo {title} {{Theory of Kinetic Phenomena in Liquid $^3$He}},\
  }}\href@noop {} {\bibfield  {journal} {\bibinfo  {journal} {Sov. Phys. JETP}\
  }\textbf {\bibinfo {volume} {5}},\ \bibinfo {pages} {887} (\bibinfo {year}
  {1957})}\BibitemShut {NoStop}%
\bibitem [{\citenamefont {Levin}\ and\ \citenamefont {Valls}(1983)}]{lev83}%
  \BibitemOpen
  \bibfield  {author} {\bibinfo {author} {\bibfnamefont {K.}~\bibnamefont
  {Levin}}\ and\ \bibinfo {author} {\bibfnamefont {O.~T.}\ \bibnamefont
  {Valls}},\ }\bibfield  {{Phenomenological Theories of Liquid $^3$He}} {\emph
  {\bibinfo {title} {{Phenomenological Theories of Liquid $^3$He}},\ }}\href
  {\doibase 10.1016/0370-1573(83)90049-2} {\bibfield  {journal} {\bibinfo
  {journal} {Phys. Rep.}\ }\textbf {\bibinfo {volume} {98}},\ \bibinfo {pages}
  {1} (\bibinfo {year} {1983})}\BibitemShut {NoStop}%
\bibitem [{\citenamefont {Doniach}\ and\ \citenamefont
  {Engelsberg}(1966)}]{don66}%
  \BibitemOpen
  \bibfield  {author} {\bibinfo {author} {\bibfnamefont {S.}~\bibnamefont
  {Doniach}}\ and\ \bibinfo {author} {\bibfnamefont {S.}~\bibnamefont
  {Engelsberg}},\ }\bibfield  {{Low-Temperature Properties of Nearly
  Ferromagnetic Fermi Liquids}} {\emph {\bibinfo {title} {{Low-Temperature
  Properties of Nearly Ferromagnetic Fermi Liquids}},\ }}\href {\doibase
  10.1103/PhysRevLett.17.750} {\bibfield  {journal} {\bibinfo  {journal} {Phys.
  Rev. Lett.}\ }\textbf {\bibinfo {volume} {17}},\ \bibinfo {pages} {750}
  (\bibinfo {year} {1966})}\BibitemShut {NoStop}%
\bibitem [{\citenamefont {Pethick}(1969)}]{pet69}%
  \BibitemOpen
  \bibfield  {author} {\bibinfo {author} {\bibfnamefont {C.~J.}\ \bibnamefont
  {Pethick}},\ }\bibinfo {title} {Quantum fluids and nuclear matter},\ \
  (\bibinfo  {publisher} {Gordon and Breach},\ \bibinfo {year} {1969})\ Chap.\
  \bibinfo {chapter} {{Selected Topics in the Theory of Normal Fermi Liquids}},
  pp.\ \bibinfo {pages} {187--251}\BibitemShut {NoStop}%
\bibitem [{\citenamefont {Layzer}\ and\ \citenamefont {Fay}(1971)}]{lay71}%
  \BibitemOpen
  \bibfield  {author} {\bibinfo {author} {\bibfnamefont {A.}~\bibnamefont
  {Layzer}}\ and\ \bibinfo {author} {\bibfnamefont {D.}~\bibnamefont {Fay}},\
  }\bibfield  {{Spin-Fluctuation Exchange Mechanism for P-wave Pairing in
  Liquid $^3$He}} {\emph {\bibinfo {title} {{Spin-Fluctuation Exchange
  Mechanism for P-wave Pairing in Liquid $^3$He}},\ }}\href@noop {} {\bibfield
  {journal} {\bibinfo  {journal} {Int. J. Magn.}\ }\textbf {\bibinfo {volume}
  {1}},\ \bibinfo {pages} {135} (\bibinfo {year} {1971})}\BibitemShut {NoStop}%
\bibitem [{\citenamefont {Layzer}\ and\ \citenamefont {Fay}(1974)}]{lay74}%
  \BibitemOpen
  \bibfield  {author} {\bibinfo {author} {\bibfnamefont {A.}~\bibnamefont
  {Layzer}}\ and\ \bibinfo {author} {\bibfnamefont {D.}~\bibnamefont {Fay}},\
  }\bibfield  {{Spin-Fluctuation Exchange: Mechanism for a Superfluid
  Transition in Liquid $^3$He}} {\emph {\bibinfo {title} {{Spin-Fluctuation
  Exchange: Mechanism for a Superfluid Transition in Liquid $^3$He}},\ }}\href
  {\doibase 10.1016/0038-1098(74)91152-1} {\bibfield  {journal} {\bibinfo
  {journal} {Sol. State Comm.}\ }\textbf {\bibinfo {volume} {15}},\ \bibinfo
  {pages} {599} (\bibinfo {year} {1974})}\BibitemShut {NoStop}%
\bibitem [{\citenamefont {Anderson}\ and\ \citenamefont
  {Brinkman}(1975)}]{and75}%
  \BibitemOpen
  \bibfield  {author} {\bibinfo {author} {\bibfnamefont {P.~W.}\ \bibnamefont
  {Anderson}}\ and\ \bibinfo {author} {\bibfnamefont {W.~F.}\ \bibnamefont
  {Brinkman}},\ }in\ \href@noop {} {\emph {\bibinfo {booktitle} {Helium
  Liquids}}},\ \bibinfo {editor} {edited by\ \bibinfo {editor} {\bibnamefont
  {edited~by J.~G. M.~Armitage}}\ and\ \bibinfo {editor} {\bibfnamefont
  {I.~E.}\ \bibnamefont {Farquhar}}}\ (\bibinfo  {publisher} {Academic Press,
  New York},\ \bibinfo {year} {1975)})\ p.\ \bibinfo {pages} {p.
  315}\BibitemShut {NoStop}%
\bibitem [{\citenamefont {Leggett}(1975)}]{leg75}%
  \BibitemOpen
  \bibfield  {author} {\bibinfo {author} {\bibfnamefont {A.~J.}\ \bibnamefont
  {Leggett}},\ }\bibfield  {{Theoretical Description of the New Phases of
  Liquid $^3$He}} {\emph {\bibinfo {title} {{Theoretical Description of the New
  Phases of Liquid $^3$He}},\ }}\href {\doibase 10.1103/RevModPhys.47.331}
  {\bibfield  {journal} {\bibinfo  {journal} {Rev. Mod. Phys.}\ }\textbf
  {\bibinfo {volume} {47}},\ \bibinfo {pages} {331} (\bibinfo {year}
  {1975})}\BibitemShut {NoStop}%
\bibitem [{\citenamefont {Vollhardt}\ and\ \citenamefont
  {W\"olfle}(1990)}]{vollhardt90}%
  \BibitemOpen
  \bibfield  {author} {\bibinfo {author} {\bibfnamefont {D.}~\bibnamefont
  {Vollhardt}}\ and\ \bibinfo {author} {\bibfnamefont {P.}~\bibnamefont
  {W\"olfle}},\ }\href@noop {} {\emph {\bibinfo {title} {{The Superfluid Phases
  of $^3$He}}}}\ (\bibinfo  {publisher} {Taylor \& Francis},\ \bibinfo
  {address} {New York},\ \bibinfo {year} {1990})\BibitemShut {NoStop}%
\bibitem [{\citenamefont {Dobbs}(2000)}]{dobbs00}%
  \BibitemOpen
  \bibfield  {author} {\bibinfo {author} {\bibfnamefont {E.~R.}\ \bibnamefont
  {Dobbs}},\ }\href@noop {} {\emph {\bibinfo {title} {Helium Three}}}\
  (\bibinfo  {publisher} {Oxford University Press},\ \bibinfo {address}
  {Oxford, England},\ \bibinfo {year} {2000})\BibitemShut {NoStop}%
\bibitem [{\citenamefont {Keldysh}(1965)}]{kel65}%
  \BibitemOpen
  \bibfield  {author} {\bibinfo {author} {\bibfnamefont {L.~V.}\ \bibnamefont
  {Keldysh}},\ }\bibfield  {{Diagram Technique for Nonequilibrium Processes}}
  {\emph {\bibinfo {title} {{Diagram Technique for Nonequilibrium Processes}},\
  }}\href@noop {} {\bibfield  {journal} {\bibinfo  {journal} {Zh. Eskp. Teor.
  Fiz.}\ }\textbf {\bibinfo {volume} {47}},\ \bibinfo {pages} {1515} (\bibinfo
  {year} {1965})},\ \bibinfo {note} {english: Sov. Phys. JETP, 20, 1018
  (1965)}\BibitemShut {NoStop}%
\bibitem [{\citenamefont {Lin}\ and\ \citenamefont {Sauls}(2021)}]{lin21}%
  \BibitemOpen
  \bibfield  {author} {\bibinfo {author} {\bibfnamefont {W.-T.}\ \bibnamefont
  {Lin}}\ and\ \bibinfo {author} {\bibfnamefont {J.~A.}\ \bibnamefont
  {Sauls}},\ }\bibfield  {{Effects of Incipient Pairing on Non-equilibrium
  Quasiparticle Transport in Fermi Liquids}} {\emph {\bibinfo {title} {{Effects
  of Incipient Pairing on Non-equilibrium Quasiparticle Transport in Fermi
  Liquids}},\ }}\href {https://arxiv.org/abs/2110.10339} {\bibfield  {journal}
  {\bibinfo  {journal} {Proc. Roy. Soc. A}\ }\textbf {\bibinfo {volume}
  {submitted}},\ \bibinfo {pages} {1} (\bibinfo {year} {2021})},\ \bibinfo
  {note} {https://arxiv.org/abs/2110.10339}\BibitemShut {NoStop}%
\bibitem [{\citenamefont {Kita}(2010)}]{kit10}%
  \BibitemOpen
  \bibfield  {author} {\bibinfo {author} {\bibfnamefont {T.}~\bibnamefont
  {Kita}},\ }\bibfield  {{Introduction to Nonequilibrium Statistical Mechanics
  with Quantum Field Theory}} {\emph {\bibinfo {title} {{Introduction to
  Nonequilibrium Statistical Mechanics with Quantum Field Theory}},\ }}\href
  {\doibase 10.1143/PTP.123.581} {\bibfield  {journal} {\bibinfo  {journal}
  {Prog. Theor. Phys.}\ }\textbf {\bibinfo {volume} {123}},\ \bibinfo {pages}
  {581} (\bibinfo {year} {2010})}\BibitemShut {NoStop}%
\bibitem [{\citenamefont {Serene}\ and\ \citenamefont {Rainer}(1983)}]{ser83}%
  \BibitemOpen
  \bibfield  {author} {\bibinfo {author} {\bibfnamefont {J.~W.}\ \bibnamefont
  {Serene}}\ and\ \bibinfo {author} {\bibfnamefont {D.}~\bibnamefont
  {Rainer}},\ }\bibfield  {{The Quasiclassical Approach to $^3He$}} {\emph
  {\bibinfo {title} {{The Quasiclassical Approach to $^3He$}},\ }}\href
  {\doibase 10.1016/0370-1573(83)90051-0} {\bibfield  {journal} {\bibinfo
  {journal} {Phys. Rep.}\ }\textbf {\bibinfo {volume} {101}},\ \bibinfo {pages}
  {221} (\bibinfo {year} {1983})}\BibitemShut {NoStop}%
\bibitem [{\citenamefont {Rainer}\ and\ \citenamefont {Sauls}(1994)}]{rai94b}%
  \BibitemOpen
  \bibfield  {author} {\bibinfo {author} {\bibfnamefont {D.}~\bibnamefont
  {Rainer}}\ and\ \bibinfo {author} {\bibfnamefont {J.~A.}\ \bibnamefont
  {Sauls}},\ }\bibinfo {title} {{Strong-Coupling Theory of
  Superconductivity}},\ in\ \href {\doibase 10.1142/9789814503891_0002} {\emph
  {\bibinfo {booktitle} {Superconductivity: From Basic Physics to New
  Developments}}}\ (\bibinfo  {publisher} {World Scientific},\ \bibinfo
  {address} {Singapore},\ \bibinfo {year} {1994})\ pp.\ \bibinfo {pages}
  {45--78},\ \bibinfo {note}
  {arXiv:https://arxiv.org/abs/1809.05264}\BibitemShut {NoStop}%
\bibitem [{\citenamefont {Abrikosov}\ \emph {et~al.}(1963)\citenamefont
  {Abrikosov}, \citenamefont {Gorkov},\ and\ \citenamefont
  {Dzyaloshinski}}]{AGD}%
  \BibitemOpen
  \bibfield  {author} {\bibinfo {author} {\bibfnamefont {A.}~\bibnamefont
  {Abrikosov}}, \bibinfo {author} {\bibfnamefont {L.}~\bibnamefont {Gorkov}}, \
  and\ \bibinfo {author} {\bibfnamefont {I.}~\bibnamefont {Dzyaloshinski}},\
  }\href@noop {} {\emph {\bibinfo {title} {``{Methods of Quantum Field Theory
  in Statistical Physics}''}}}\ (\bibinfo  {publisher} {Prentice-Hall, Inc.},\
  \bibinfo {address} {Englewood Cliffs, NJ},\ \bibinfo {year}
  {1963})\BibitemShut {NoStop}%
\bibitem [{Note1()}]{Note1}%
  \BibitemOpen
  \bibinfo {note} {Here we consider non-magnetic disturbances of the
  quasiparticle distribution in which case only the spin-independent Landau
  interaction is relevant and the sum over spin states contributes a factor of
  $2$.}\BibitemShut {Stop}%
\bibitem [{\citenamefont {Brooker}\ and\ \citenamefont {Sykes}(1970)}]{bro70}%
  \BibitemOpen
  \bibfield  {author} {\bibinfo {author} {\bibfnamefont {G.~A.}\ \bibnamefont
  {Brooker}}\ and\ \bibinfo {author} {\bibfnamefont {J.}~\bibnamefont
  {Sykes}},\ }\bibfield  {{Sound Propagation and Relaxation Times of a Fermi
  Liquid}} {\emph {\bibinfo {title} {{Sound Propagation and Relaxation Times of
  a Fermi Liquid}},\ }}\href {\doibase 10.1016/0003-4916(70)90290-3} {\bibfield
   {journal} {\bibinfo  {journal} {Ann. Phys.}\ }\textbf {\bibinfo {volume}
  {61}},\ \bibinfo {pages} {387} (\bibinfo {year} {1970})}\BibitemShut
  {NoStop}%
\bibitem [{\citenamefont {Abrikosov}(1978)}]{abrikosov78}%
  \BibitemOpen
  \bibfield  {author} {\bibinfo {author} {\bibfnamefont {A.~A.}\ \bibnamefont
  {Abrikosov}},\ }\bibinfo {title} {{Theory of Normal Metals}},\ \ (\bibinfo
  {publisher} {Wiley Publishers},\ \bibinfo {address} {New York},\ \bibinfo
  {year} {1978})\BibitemShut {NoStop}%
\bibitem [{\citenamefont {Baym}\ and\ \citenamefont {Pethick}(1978)}]{bay78}%
  \BibitemOpen
  \bibfield  {author} {\bibinfo {author} {\bibfnamefont {G.}~\bibnamefont
  {Baym}}\ and\ \bibinfo {author} {\bibfnamefont {C.~J.}\ \bibnamefont
  {Pethick}},\ }\bibinfo {title} {{Landau Fermi Liquid Theory and the Low
  Temperature Properties of Liquid $^3$He}},\ in\ \href@noop {} {\emph
  {\bibinfo {booktitle} {{The Physics of Solid and Liquid Helium, Part 2}}}},\
  \bibinfo {editor} {edited by\ \bibinfo {editor} {\bibfnamefont {K.~H.}\
  \bibnamefont {Benneman}}\ and\ \bibinfo {editor} {\bibfnamefont {J.~B.}\
  \bibnamefont {Ketterson}}}\ (\bibinfo  {publisher} {Wiley, New York},\
  \bibinfo {year} {1978})\ pp.\ \bibinfo {pages} {1--122}\BibitemShut {NoStop}%
\bibitem [{\citenamefont {Sauls}\ and\ \citenamefont {Sharma}(2010)}]{sau10}%
  \BibitemOpen
  \bibfield  {author} {\bibinfo {author} {\bibfnamefont {J.~A.}\ \bibnamefont
  {Sauls}}\ and\ \bibinfo {author} {\bibfnamefont {P.}~\bibnamefont {Sharma}},\
  }\bibfield  {{Theory of Heat Transport of Normal Liquid $^3$He in Aerogel}}
  {\emph {\bibinfo {title} {{Theory of Heat Transport of Normal Liquid $^3$He
  in Aerogel}},\ }}\href {\doibase 10.1088/1367-2630/12/8/083056} {\bibfield
  {journal} {\bibinfo  {journal} {New Journal of Physics}\ }\textbf {\bibinfo
  {volume} {12}},\ \bibinfo {pages} {083056} (\bibinfo {year}
  {2010})}\BibitemShut {NoStop}%
\bibitem [{\citenamefont {Sauls}\ and\ \citenamefont {Serene}(1981)}]{sau81b}%
  \BibitemOpen
  \bibfield  {author} {\bibinfo {author} {\bibfnamefont {J.~A.}\ \bibnamefont
  {Sauls}}\ and\ \bibinfo {author} {\bibfnamefont {J.~W.}\ \bibnamefont
  {Serene}},\ }\bibfield  {{Potential Scattering Models for the Quasiparticle
  Interactions in Liquid $^3$He}} {\emph {\bibinfo {title} {{Potential
  Scattering Models for the Quasiparticle Interactions in Liquid $^3$He}},\
  }}\href {\doibase 10.1103/PhysRevB.24.183} {\bibfield  {journal} {\bibinfo
  {journal} {Phys. Rev. B}\ }\textbf {\bibinfo {volume} {24}},\ \bibinfo
  {pages} {183} (\bibinfo {year} {1981})}\BibitemShut {NoStop}%
\bibitem [{\citenamefont {Paulson}\ and\ \citenamefont
  {Wheatley}(1978)}]{pau78c}%
  \BibitemOpen
  \bibfield  {author} {\bibinfo {author} {\bibfnamefont {D.~N.}\ \bibnamefont
  {Paulson}}\ and\ \bibinfo {author} {\bibfnamefont {J.~C.}\ \bibnamefont
  {Wheatley}},\ }\bibfield  {{Incipient Superfluidity in Liquid $^3$He above
  the Superfluid Transition Temperature}} {\emph {\bibinfo {title} {{Incipient
  Superfluidity in Liquid $^3$He above the Superfluid Transition
  Temperature}},\ }}\href {\doibase 10.1103/PhysRevLett.41.561} {\bibfield
  {journal} {\bibinfo  {journal} {Phys. Rev. Lett.}\ }\textbf {\bibinfo
  {volume} {41}},\ \bibinfo {pages} {561} (\bibinfo {year} {1978})}\BibitemShut
  {NoStop}%
\bibitem [{\citenamefont {Emery}(1976)}]{eme76}%
  \BibitemOpen
  \bibfield  {author} {\bibinfo {author} {\bibfnamefont {V.~J.}\ \bibnamefont
  {Emery}},\ }\bibfield  {{Fluctuations above the superfluid transition in
  liquid $^3$He}} {\emph {\bibinfo {title} {{Fluctuations above the superfluid
  transition in liquid $^3$He}},\ }}\href {\doibase 10.1007/BF00654719}
  {\bibfield  {journal} {\bibinfo  {journal} {J. Low Temp. Phys.}\ }\textbf
  {\bibinfo {volume} {22}},\ \bibinfo {pages} {467} (\bibinfo {year}
  {1976})}\BibitemShut {NoStop}%
\bibitem [{\citenamefont {Samalam}\ and\ \citenamefont {Serene}(1978)}]{sam78}%
  \BibitemOpen
  \bibfield  {author} {\bibinfo {author} {\bibfnamefont {V.~K.}\ \bibnamefont
  {Samalam}}\ and\ \bibinfo {author} {\bibfnamefont {J.~W.}\ \bibnamefont
  {Serene}},\ }\bibfield  {{Zero-Sound Attenuation from Order-Parameter
  Fluctuations in Liquid $^3$He}} {\emph {\bibinfo {title} {{Zero-Sound
  Attenuation from Order-Parameter Fluctuations in Liquid $^3$He}},\ }}\href
  {\doibase 10.1103/PhysRevLett.41.497} {\bibfield  {journal} {\bibinfo
  {journal} {Phys. Rev. Lett.}\ }\textbf {\bibinfo {volume} {41}},\ \bibinfo
  {pages} {497} (\bibinfo {year} {1978})}\BibitemShut {NoStop}%
\bibitem [{\citenamefont {Halperin}\ and\ \citenamefont
  {Varoquaux}(1990)}]{hal90}%
  \BibitemOpen
  \bibfield  {author} {\bibinfo {author} {\bibfnamefont {W.~P.}\ \bibnamefont
  {Halperin}}\ and\ \bibinfo {author} {\bibfnamefont {E.}~\bibnamefont
  {Varoquaux}},\ }\bibinfo {title} {Order {P}arameter {C}ollective {M}odes in
  {S}uperfluid {{$^3He$}}},\ in\ \href@noop {} {\emph {\bibinfo {booktitle}
  {Helium Three}}},\ \bibinfo {editor} {edited by\ \bibinfo {editor}
  {\bibfnamefont {W.~P.}\ \bibnamefont {Halperin}}\ and\ \bibinfo {editor}
  {\bibfnamefont {L.~P.}\ \bibnamefont {Pitaevskii}}}\ (\bibinfo  {publisher}
  {Elsevier Science Publishers},\ \bibinfo {address} {Amsterdam},\ \bibinfo
  {year} {1990})\ p.\ \bibinfo {pages} {353}\BibitemShut {NoStop}%
\bibitem [{\citenamefont {Corruccini}\ \emph {et~al.}(1969)\citenamefont
  {Corruccini}, \citenamefont {Clark}, \citenamefont {Mermin},\ and\
  \citenamefont {Wilkins}}]{cor69}%
  \BibitemOpen
  \bibfield  {author} {\bibinfo {author} {\bibfnamefont {L.}~\bibnamefont
  {Corruccini}}, \bibinfo {author} {\bibfnamefont {J.~S.}\ \bibnamefont
  {Clark}}, \bibinfo {author} {\bibfnamefont {N.~D.}\ \bibnamefont {Mermin}}, \
  and\ \bibinfo {author} {\bibfnamefont {J.~W.}\ \bibnamefont {Wilkins}},\
  }\bibfield  {{Attenuation of Transverse Zero Sound in $^3$He}} {\emph
  {\bibinfo {title} {{Attenuation of Transverse Zero Sound in $^3$He}},\
  }}\href {\doibase 10.1103/PhysRev.180.225} {\bibfield  {journal} {\bibinfo
  {journal} {Phys. Rev.}\ }\textbf {\bibinfo {volume} {180}},\ \bibinfo {pages}
  {225} (\bibinfo {year} {1969})}\BibitemShut {NoStop}%
\bibitem [{\citenamefont {Abel}\ \emph {et~al.}(1966)\citenamefont {Abel},
  \citenamefont {Anderson},\ and\ \citenamefont {Wheatley}}]{abe66}%
  \BibitemOpen
  \bibfield  {author} {\bibinfo {author} {\bibfnamefont {W.~R.}\ \bibnamefont
  {Abel}}, \bibinfo {author} {\bibfnamefont {A.~C.}\ \bibnamefont {Anderson}},
  \ and\ \bibinfo {author} {\bibfnamefont {J.~C.}\ \bibnamefont {Wheatley}},\
  }\bibfield  {{Propagation of Zero Sound in Liquid $^{3}$He at Low
  Temperatures}} {\emph {\bibinfo {title} {{Propagation of Zero Sound in Liquid
  $^{3}$He at Low Temperatures}},\ }}\href {\doibase 10.1103/PhysRevLett.17.74}
  {\bibfield  {journal} {\bibinfo  {journal} {Phys. Rev. Lett.}\ }\textbf
  {\bibinfo {volume} {17}},\ \bibinfo {pages} {74} (\bibinfo {year}
  {1966})}\BibitemShut {NoStop}%
\bibitem [{Note2()}]{Note2}%
  \BibitemOpen
  \bibinfo {note} {Emery's formula would correspond to a diagram similar to
  Fig.~\ref {fmf-Self_Energies}(c), but with $\Gamma \rightarrow \Gamma _Q$
  \protect \emph {twice}, and the internal particle-hole pair replaced by a
  particle-particle pair, thus double counting the particle-particle ladder
  diagrams, which explains why Emery's expression also has a prefactor a factor
  of 2 larger than our prefactor.}\BibitemShut {Stop}%
\bibitem [{Note3()}]{Note3}%
  \BibitemOpen
  \bibinfo {note} {See Supplemental Material below for details of the analysis
  supporting the corrections to the transport equation, application of the
  theory to the anomalous attenuation of zero sound for $T\rightarrow T_{c}^+$,
  and comparison with the cutoff procedure of Samalam and Serene.~\cite
  {sam78}}\BibitemShut {NoStop}%
\end{thebibliography}

\begin{thebibliography}{6}%
\makeatletter
\providecommand \@ifxundefined [1]{%
 \@ifx{#1\undefined}
}%
\providecommand \@ifnum [1]{%
 \ifnum #1\expandafter \@firstoftwo
 \else \expandafter \@secondoftwo
 \fi
}%
\providecommand \@ifx [1]{%
 \ifx #1\expandafter \@firstoftwo
 \else \expandafter \@secondoftwo
 \fi
}%
\providecommand \natexlab [1]{#1}%
\providecommand \enquote  [1]{``#1''}%
\providecommand \bibnamefont  [1]{#1}%
\providecommand \bibfnamefont [1]{#1}%
\providecommand \citenamefont [1]{#1}%
\providecommand \href@noop [0]{\@secondoftwo}%
\providecommand \href [0]{\begingroup \@sanitize@url \@href}%
\providecommand \@href[1]{\@@startlink{#1}\@@href}%
\providecommand \@@href[1]{\endgroup#1\@@endlink}%
\providecommand \@sanitize@url [0]{\catcode `\\12\catcode `\$12\catcode
  `\&12\catcode `\#12\catcode `\^12\catcode `\_12\catcode `\%12\relax}%
\providecommand \@@startlink[1]{}%
\providecommand \@@endlink[0]{}%
\providecommand \url  [0]{\begingroup\@sanitize@url \@url }%
\providecommand \@url [1]{\endgroup\@href {#1}{\urlprefix }}%
\providecommand \urlprefix  [0]{URL }%
\providecommand \Eprint [0]{\href }%
\providecommand \doibase [0]{http://dx.doi.org/}%
\providecommand \selectlanguage [0]{\@gobble}%
\providecommand \bibinfo  [0]{\@secondoftwo}%
\providecommand \bibfield  [0]{\@secondoftwo}%
\providecommand \translation [1]{[#1]}%
\providecommand \BibitemOpen [0]{}%
\providecommand \bibitemStop [0]{}%
\providecommand \bibitemNoStop [0]{.\EOS\space}%
\providecommand \EOS [0]{\spacefactor3000\relax}%
\providecommand \BibitemShut  [1]{\csname bibitem#1\endcsname}%
\let\auto@bib@innerbib\@empty
\bibitem [{\citenamefont {Corruccini}\ \emph {et~al.}(1969)\citenamefont
  {Corruccini}, \citenamefont {Clark}, \citenamefont {Mermin},\ and\
  \citenamefont {Wilkins}}]{cor69s}%
  \BibitemOpen
  \bibfield  {author} {\bibinfo {author} {\bibfnamefont {L.}~\bibnamefont
  {Corruccini}}, \bibinfo {author} {\bibfnamefont {J.~S.}\ \bibnamefont
  {Clark}}, \bibinfo {author} {\bibfnamefont {N.~D.}\ \bibnamefont {Mermin}}, \
  and\ \bibinfo {author} {\bibfnamefont {J.~W.}\ \bibnamefont {Wilkins}},\
  }\bibfield  {{Attenuation of Transverse Zero Sound in $^3$He}} {\emph
  {\bibinfo {title} {{Attenuation of Transverse Zero Sound in $^3$He}},\
  }}\href {\doibase 10.1103/PhysRev.180.225} {\bibfield  {journal} {\bibinfo
  {journal} {Phys. Rev.}\ }\textbf {\bibinfo {volume} {180}},\ \bibinfo {pages}
  {225} (\bibinfo {year} {1969})}\BibitemShut {NoStop}%
\bibitem [{\citenamefont {Baym}\ and\ \citenamefont {Pethick}(1978)}]{bay78s}%
  \BibitemOpen
  \bibfield  {author} {\bibinfo {author} {\bibfnamefont {G.}~\bibnamefont
  {Baym}}\ and\ \bibinfo {author} {\bibfnamefont {C.~J.}\ \bibnamefont
  {Pethick}},\ }\bibinfo {title} {{Landau Fermi Liquid Theory and the Low
  Temperature Properties of Liquid $^3$He}},\ in\ \href@noop {} {\emph
  {\bibinfo {booktitle} {{The Physics of Solid and Liquid Helium, Part 2}}}},\
  \bibinfo {editor} {edited by\ \bibinfo {editor} {\bibfnamefont {K.~H.}\
  \bibnamefont {Benneman}}\ and\ \bibinfo {editor} {\bibfnamefont {J.~B.}\
  \bibnamefont {Ketterson}}}\ (\bibinfo  {publisher} {Wiley, New York},\
  \bibinfo {year} {1978})\ pp.\ \bibinfo {pages} {1--122}\BibitemShut {NoStop}%
\bibitem [{\citenamefont {Paulson}\ and\ \citenamefont
  {Wheatley}(1978)}]{pau78s}%
  \BibitemOpen
  \bibfield  {author} {\bibinfo {author} {\bibfnamefont {D.~N.}\ \bibnamefont
  {Paulson}}\ and\ \bibinfo {author} {\bibfnamefont {J.~C.}\ \bibnamefont
  {Wheatley}},\ }\bibfield  {{Incipient Superfluidity in Liquid $^3$He above
  the Superfluid Transition Temperature}} {\emph {\bibinfo {title} {{Incipient
  Superfluidity in Liquid $^3$He above the Superfluid Transition
  Temperature}},\ }}\href {\doibase 10.1103/PhysRevLett.41.561} {\bibfield
  {journal} {\bibinfo  {journal} {Phys. Rev. Lett.}\ }\textbf {\bibinfo
  {volume} {41}},\ \bibinfo {pages} {561} (\bibinfo {year} {1978})}\BibitemShut
  {NoStop}%
\bibitem [{\citenamefont {Haard}(2000)}]{har00s}%
  \BibitemOpen
  \bibfield  {author} {\bibinfo {author} {\bibfnamefont {T.}~\bibnamefont
  {Haard}},\ }\bibfield  {{$^3$He Calculator}} {\emph {\bibinfo {title}
  {{$^3$He Calculator}},\ }}\href {\doibase
  http://spindry.phys.northwestern.edu/he3.htm} {\bibfield  {journal} {\bibinfo
   {journal} {{ULT@Northwestern}}\ }\textbf {\bibinfo {volume} {{Online}}}
  (\bibinfo {year} {2000}),\
  http://spindry.phys.northwestern.edu/he3.htm}\BibitemShut {NoStop}%
\bibitem [{\citenamefont {Samalam}\ and\ \citenamefont {Serene}(1978)}]{sam78s}%
  \BibitemOpen
  \bibfield  {author} {\bibinfo {author} {\bibfnamefont {V.~K.}\ \bibnamefont
  {Samalam}}\ and\ \bibinfo {author} {\bibfnamefont {J.~W.}\ \bibnamefont
  {Serene}},\ }\bibfield  {{Zero-Sound Attenuation from Order-Parameter
  Fluctuations in Liquid $^3$He}} {\emph {\bibinfo {title} {{Zero-Sound
  Attenuation from Order-Parameter Fluctuations in Liquid $^3$He}},\ }}\href
  {\doibase 10.1103/PhysRevLett.41.497} {\bibfield  {journal} {\bibinfo
  {journal} {Phys. Rev. Lett.}\ }\textbf {\bibinfo {volume} {41}},\ \bibinfo
  {pages} {497} (\bibinfo {year} {1978})}\BibitemShut {NoStop}%
\bibitem [{\citenamefont {Dobbs}(2000)}]{dobbs00s}%
  \BibitemOpen
  \bibfield  {author} {\bibinfo {author} {\bibfnamefont {E.~R.}\ \bibnamefont
  {Dobbs}},\ }\href@noop {} {\emph {\bibinfo {title} {Helium Three}}}\
  (\bibinfo  {publisher} {Oxford University Press},\ \bibinfo {address}
  {Oxford, England},\ \bibinfo {year} {2000})\BibitemShut {NoStop}%
\end{thebibliography}
%

\end{document}